\documentclass[usenatbib]{mn2e}

\usepackage[fleqn]{amsmath} 
\usepackage{txfonts}
\usepackage{float}
\usepackage{natbib}
\usepackage{amssymb}
\usepackage{journals}
\usepackage{graphicx}
\usepackage{booktabs}
\usepackage{color}

\newcommand{\machnum}{\mathcal{M}}

\newcommand{\myvec}{\boldsymbol}

\newcommand{\mynabla}{\myvec{\nabla}}

\newcommand{\mastervar}{temperature }

\newcommand{\mytilde}{\raise.17ex\hbox{$\scriptstyle\sim$}}

\title[Shock statistics in non-radiative moving-mesh
simulations]{Shock finding on a moving-mesh: I. Shock statistics in
  non-radiative cosmological simulations}

\author[K.~Schaal and V.~Springel]
  {Kevin~Schaal$^{1,2}$\thanks{e-mail: kevin.schaal@h-its.org;
  volker.springel@h-its.org} and
  Volker~Springel$^{1,2}$\footnotemark[1] \\ 
  $^1$Heidelberg Institute for Theoretical Studies, Schloss-Wolfsbrunnenweg 35, 69118 Heidelberg, Germany\\
  $^2$Zentrum f\"ur Astronomie der Universit\"at Heidelberg,
  Astronomisches Recheninstitut, M\"{o}nchhofstr. 12-14, 69120
  Heidelberg, Germany}

\begin{document}

\pagerange{\pageref{firstpage}--\pageref{lastpage}} \pubyear{2015} 

\maketitle

\label{firstpage}

\begin{abstract}
  Cosmological shock waves play an important role in hierarchical
  structure formation by dissipating and thermalizing kinetic energy
  of gas flows, thereby heating the Universe. Furthermore, identifying
  shocks in hydrodynamical simulations and measuring their Mach number
  accurately are critical for calculating the production of non-thermal
  particle components through diffusive shock acceleration. However,
  shocks are often significantly broadened in numerical simulations,
  making it challenging to implement an accurate shock finder.  We
  here introduce a refined methodology for detecting shocks in the
  moving-mesh code {\small AREPO}, and show that results for shock statistics
  can be sensitive to implementation details. We put special emphasis
  on filtering against spurious shock detections due to tangential
  discontinuities and contacts. Both of them are omnipresent in
  cosmological simulations, for example in the form of shear-induced
  Kelvin-Helmholtz instabilities and cold fronts. As an initial
  application of our new implementation, we analyse shock statistics
  in non-radiative cosmological simulations of dark matter and
  baryons.  We find that the bulk of energy dissipation at redshift
  zero occurs in shocks with Mach numbers around $\machnum\approx
  2.7$. Furthermore, almost $40\%$ of the thermalization is
  contributed by shocks in the warm hot intergalactic medium,
  whereas $\approx60\%$ occurs in clusters, groups, and smaller haloes.
  Compared to previous studies, these findings revise the
  characterization of the most important shocks towards higher Mach
  numbers and lower density structures.  Our results also suggest that
  regions with densities above and below $\delta_\mathrm{b}=100$ should be
  roughly equally important for the energetics of cosmic ray
  acceleration through large-scale structure shocks.
\end{abstract}


\section{Introduction}

The collapse of dark and baryonic matter during hierarchical
large-scale structure formation releases gravitational energy and
transforms it into kinetic energy.  The bulk of the kinetic energy of
the gas gets dissipated by cosmological shocks, heating the gas in
virialized haloes (e.g the intracluster medium, ICM) as well as
in the warm-hot intergalactic medium (WHIM). Cosmological hydrodynamic
shocks are collisionless; they are established due to plasma
interactions by means of magnetic fields. They can themselves amplify
magnetic fields and accelerate particles via diffusive shock
acceleration \citep[DSA;][]{AXFORD_1977, KRYMSKI_1977, BELL_1978_1, 
BELL_1978_2, BLANDFORD, MALKOVREVIEW} up to relativistic
energies, producing cosmic rays.

Directly observing cosmological shocks is challenging, especially
outside cluster cores where the X-ray emission is weak.  An obvious
approach is to look for jumps in the thermal gas quantities.  In this
way, and using exquisite X-ray data from the \textit{Chandra}
telescope, the first merger shocks have been confirmed in the bullet
cluster \citep[$\machnum \approx 3$;][]{MARKEVITCH2002,MARKEVITCH2006}
and in the train-wreck cluster \citep[$\machnum \approx
2.1$;][]{MARKEVITCH_TRAIN}. Maps of the gas density and the temperature
can be inferred from the luminosity and the spectrum of the X-ray
radiation, respectively. Both are necessary in order to calculate a
pressure map and confirm a shock.  Furthermore, it is possible to
directly measure a pressure jump by means of the thermal
Sunyaev--Zel'dovich signal \citep{SZ}. For example, steep pressure gradients have
been detected inside $R_{500}$ in the nearby Coma cluster
\citep{PLANCK_COMA}.  The location of the gradients coincides with
temperature jumps, and two shocks with Mach numbers around $\machnum\approx
2$ were reported in this way.

Shocks can also be observed indirectly at radio
wavelengths. 
Diffusively shock-accelerated cosmic ray electrons in
merger and accretion shocks produce synchrotron radiation, so-called
radio gischt \citep{ENSSLIN98, BATTAGLIA2009, PINZKE}. This phenomenon 
has been observed in several clusters \citep[see e.g.][for a review of
shocks in cluster outskirts and the associated features]{CLARKE,
  BONAFEDE, VANWEEREN, BRUEGGEN_REVIEW}.  Another radio source
triggered by shocks is the radio phoenix \citep{ENSSLIN2001_PHOENIX,
  ENSSLIN2002_PHOENIX}. In this scenario, a shock overruns fossil
radio plasma initially produced by an active galactic nucleus (AGN),
compressing the plasma and reviving its radio emission.  A radio
phoenix can reveal large-scale accretion shocks and has been reported
for the Perseus Cluster \citep{PFROMMER_PERSEUS}.

The motivation for observing shocks is manifold.  First of
all, supersonic flows and their associated shocks allow the study of
thermalization patterns and energetics of phenomena at a broad range
of spatial scales. This includes accretion shocks onto clusters,
mergers of galaxies and galaxy clusters, winds and jets of AGN, 
as well as stellar winds and supernovae.  Secondly,
observations of supernova remnants provide evidence for the creation
of non-thermal cosmic ray particles at these locations. Cosmic ray
protons can collide with thermal protons of the interstellar medium
producing pions, which subsequently decay and release
$\gamma$-radiation.  The pion decay and hence the acceleration of
cosmic ray protons has been confirmed for several supernova remnants
\citep[e.g.][]{GIULIANI_2011, ACKERMANN_2013}.

DSA and associated processes such as the modification of the
shock structure due to cosmic ray back-reaction or magnetic field
amplification have been investigated analytically
\citep[e.g.][]{DRURY_1983, MALKOV_1997, BLASI_2002, AMATO_2006}, as
well as with numerical simulations \citep[e.g.][]{ELLISON_1996,
VLADIMIROV_2006, KANG_2007, KANG_2013, FERRAND_2014}.
Furthermore, DSA can be simulated bottom-up by resolving the micro
physics with particle-in-cell (PIC) methods
\citep[e.g.][]{AMANO_2007, AMANO_2010,
RIQUELME_2011}. Alternatively, less costly hybrid methods
\citep[e.g.][]{QUEST_1988, CAPRIOLI_2014} can be used, where the
ions are treated kinetically and the electrons are modelled as a
fluid.  While basic predictions of radio, X-ray and $\gamma$-ray
emission of DSA models can be confirmed by observations of supernova
remnants \citep[e.g.][]{REYNOLDS_2008, EDMON_2011}, the
detailed understanding of the non-linear acceleration mechanism
requires further analytic and numerical work.

Additional insights will be provided by forthcoming
observations with, for example, the Cherenkov Telescope Array
\citep[CTA;][]{CTA_2011} and the Square Kilometre Array (SKA). With
the CTA it will be possible to study particle acceleration over larger
energy ranges and with increased resolution compared to present
observations.  The SKA will presumably allow a detailed study of the
magnetic field of galactic supernova remnants utilizing the effect
of Faraday rotation. Furthermore, large-scale cosmological shocks
are expected to be observed due to their synchrotron emission
\citep{KESHET_2004}.

Cosmological shocks in numerical simulations of large-scale structure
formation were analysed comprehensively in previous studies, for
example in \citet{QUILIS}, \citet{MINIATI}, \citet{RYU}, \citet{PFROMMER}, \citet{KANG},
\citet{SKILLMAN}, \citet{VAZZA}, \citet{PLANELLES}, and \citet{HONG}.  
The detected shocks can be divided into two distinct classes, external and 
internal shocks \citep{RYU}.  Strong external shocks form when previously cold and
unshocked gas ($T\lesssim 10^4$) accretes from voids onto the cosmic
web.  They have typically high Mach numbers up to $\machnum\approx
100$, but dissipate comparatively little energy due to the low
pre-shock density and temperature. Internal shocks on the other hand
occur if previously shocked and thus hotter gas inside non-linear
structures gets shock-heated further. Because of the smaller
temperature ratios compared to external shocks, the Mach numbers of
internal shocks are typically smaller ($\machnum \lesssim 10$). The
pre-shock density and temperature of internal shocks are however high.
This allows them to account for the bulk of the energy dissipation,
especially shocks with Mach numbers in the range
$2\lesssim\machnum\lesssim 4$ contribute strongly.

A detailed characterization of the prevalence and strength of shocks
in numerical simulations requires the implementation of an accurate
shock finder.  The first approaches in grid-based cosmological codes
simply used the jump conditions on a cell-by-cell basis to identify
shocked cells \citep{QUILIS, MINIATI}. As a first improvement,
\citet{RYU} proposed a method in which the shock centres are
identified in a two-step procedure. First, cells are considered to be
in a shock zone if they simultaneously meet three different criteria
meant to identify cells with some numerical shock dissipation. Within
this zone, the shock centres were then determined by looking for the
cells with the maximum compression. This more elaborate approach takes
into account that the common numerical methods capture a shock
discontinuity over a few cells, rather than exposing the full jump
strength at a single cell interface.

In order to deal with three dimensional simulations, \citet{RYU}
calculated three different Mach numbers ($\machnum_x, \machnum_y,
\machnum_z$) for each cell in the shock centre by evaluating the
temperature jump across the shock zone in each coordinate
direction. The maximum occurring Mach number was then assigned to the
shock cell.  A refinement to this method is to calculate the Mach
number via $\machnum=(\machnum_x^2+\machnum_y^2+\machnum_z^2)^{1/2}$,
thus minimizing projection effects \citep{VAZZA}.  Furthermore,
\citet{VAZZA} showed that by using the velocity jump instead of the
temperature jump slightly less scatter in the calculation was achieved
with their code.  The use of coordinate-splitting can be avoided by
characterizing the direction of shock propagation with the local
temperature gradient \citep{SKILLMAN}. In this way, a single Mach
number can be calculated and the result of the shock finder becomes
independent of the orientation between the shock and the underlying
grid. The shock-finding implementation of \citet{SKILLMAN}
additionally filters tangential discontinuities and contacts by
evaluating the pre- and post-shock temperature and density.

Quite different shock detection methodologies have been developed for
Lagrangian smoothed-particle hydrodynamics (SPH) codes. To this end,
\citet{KESHET} measured the entropy increase of each particle between different snapshots, 
and \citet{PFROMMER} measured the entropy injection rate on a per-particle
basis during the simulation. As the entropy production is directly
sourced by the artificial viscosity used for shock capturing in SPH,
this allows an estimate of the Mach number of a shock.  In another SPH
shock-finding method, \citet{HOEFT} proposed to use the local entropy
gradient for determining associated pre- and post-shock regions, and
then to calculate the Mach number across the associated jump.

In a recent code comparison project, \citet{VAZZACOMPARE} reported
reasonable agreement of different codes with respect to energy
dissipation and shock abundance as a function of Mach number.
However, significant differences have also been detected.  Especially
the detailed comparison of grid-based shock finders with the SPH-based
techniques revealed some apparent inconsistencies in the shock
morphologies and in various features in the gas phase-space diagrams.
These discrepancies in the results of the different shock finder
implementations highlight the computational challenges involved in
accurate numerical shock detection.  As we will demonstrate in this
work, a shock finder can be very sensitive to implementation details,
and it is hence crucial to improve these methods further, for example
by more carefully removing false positive shock detections associated
with tangential and contact discontinuities.

This is the goal of this paper, which has the following structure.  We
describe and validate our new methodology for finding shocks in the
moving-mesh code {\small AREPO} in Sections~\ref{sec:methodology} and
\ref{sec:validation}, respectively.  The shock finder is then applied
to non-radiative simulations in Section~\ref{sec:non_radiative}, and
differences to previous studies are discussed in
Section~\ref{sec:variations}.  Finally, we summarize our results in
Section~\ref{sec:summary}.

\section{Methodology}
\label{sec:methodology}

\subsection{The moving-mesh code {\small AREPO}}

The non-radiative cosmological simulations analysed in this paper and
the development of the shock detection method were carried out
using the {\small AREPO} code \citep{AREPO}.  In this cosmological
hydrodynamical code, the gas physics is calculated on a moving Voronoi
mesh. The mesh generating points are advected with the local velocity
of the fluid in order to achieve quasi-Lagrangian behaviour.  For
solving the Euler equations on the unstructured Voronoi grid, a finite
volume method is used in the form of a second-order unsplit Godunov
scheme with an exact Riemann solver.  With this approach the accuracy
of a grid code can be combined with features of Lagrangian codes such
as Galilean invariance and approximately constant mass per resolution
element.  Gravity exerted by the gas and the dark matter is computed
with a Tree-PM method \citep{XU, GADGET} in which long-range
gravitational forces are calculated with a particle-mesh scheme,
whereas short-range interactions are calculated in real space using a
hierarchical multipole expansion organized with an octree
\citep{BARNES}.

\subsection{The Rankine--Hugoniot jump conditions}

It is well known that the mass, momentum, and energy flux are continuous
across a discontinuity in an ideal gas. If the mass flux happens to be
zero, it follows that the normal component of the velocity and the
pressure do not jump across the discontinuity (`tangential
discontinuities'). If additionally the tangential velocity is also
continuous, a special discontinuity is present which is called a
contact.  A non-zero mass flux on the other hand implies that the
tangential velocities are continuous. In this case, a shock is present
and the normal velocities as well as the other thermodynamic variables
jump according to \citep{LL}
\begin{align}
&\frac{\rho_2}{\rho_1}=\frac{v_1}{v_2}=\frac{(\gamma+1)\machnum^2}{(\gamma-1)\machnum^2+2}, \label{eq:rho_jump}\\
&\frac{p_2}{p_1}=\frac{2\gamma\machnum^2}{\gamma+1}-\frac{\gamma-1}{\gamma+1}, \label{eq:p_jump}\\
&\frac{T_2}{T_1}=\frac{[2\gamma\machnum^2-(\gamma-1)][(\gamma-1)\machnum^2+2]}{(\gamma+1)^2\machnum^2}, \label{eq:t_jump}\\
&\frac{S_2}{S_1}=\left(\frac{2\gamma\machnum^2}{\gamma+1}-\frac{\gamma-1}{\gamma+1}\right)\left(\frac{(\gamma-1)\machnum^2+2}{(\gamma+1)\machnum^2}\right)^\gamma. \label{eq:s_jump}
\end{align}
The quantities $\rho$, $v$, $p$, $T$, and $S={p}/{\rho^\gamma}$ denote
density, velocity in the shock frame, pressure, temperature, 
and the entropic function, respectively. The Mach number $\machnum$ 
is the shock speed in the frame of the pre-shock gas, in units of the 
pre-shock sound speed $c_1$. The indices 1 and 2 label the pre-shock and 
post-shock regions, respectively, and $\gamma$ is the adiabatic index 
of the gas.

\subsection{Shock-finding method for {\small AREPO}}

We base the implementation of our shock finder on a number of previous
ideas \citep{RYU, SKILLMAN, HONG}, augmented with some
improvements. First of all, a shock zone is identified by appropriate
criteria which put special emphasis on filtering spurious shocks such
as tangential discontinuities and contacts.  We then tag cells with
maximum compression along the shock direction and inside the shock
zone as shock surface cells. The Mach number for these cells is
calculated with the temperature jump across the shock zone. Finally,
we take care of overlapping shock zones which can be present in the
case of colliding shocks.

\subsubsection{Shock direction}

For our method, the direction of shock propagation in each Voronoi
cell has to be specified.  In order to be consistent with the Mach
number calculation (see Section~\ref{sec:mach_number_calculation}), 
we use the unlimited \mastervar gradient for calculating the shock
direction:
\begin{align}
\myvec{d}_\mathrm{s}=-\frac{\mynabla T}{\left|\mynabla T\right|},
\end{align}
where $\mynabla T$ is computed with the second-order accurate
gradient operator available in {\small AREPO} for Voronoi meshes.

\subsubsection{Shock zone}
\label{sec:shock_zone}

\begin{figure}
\centering
\includegraphics{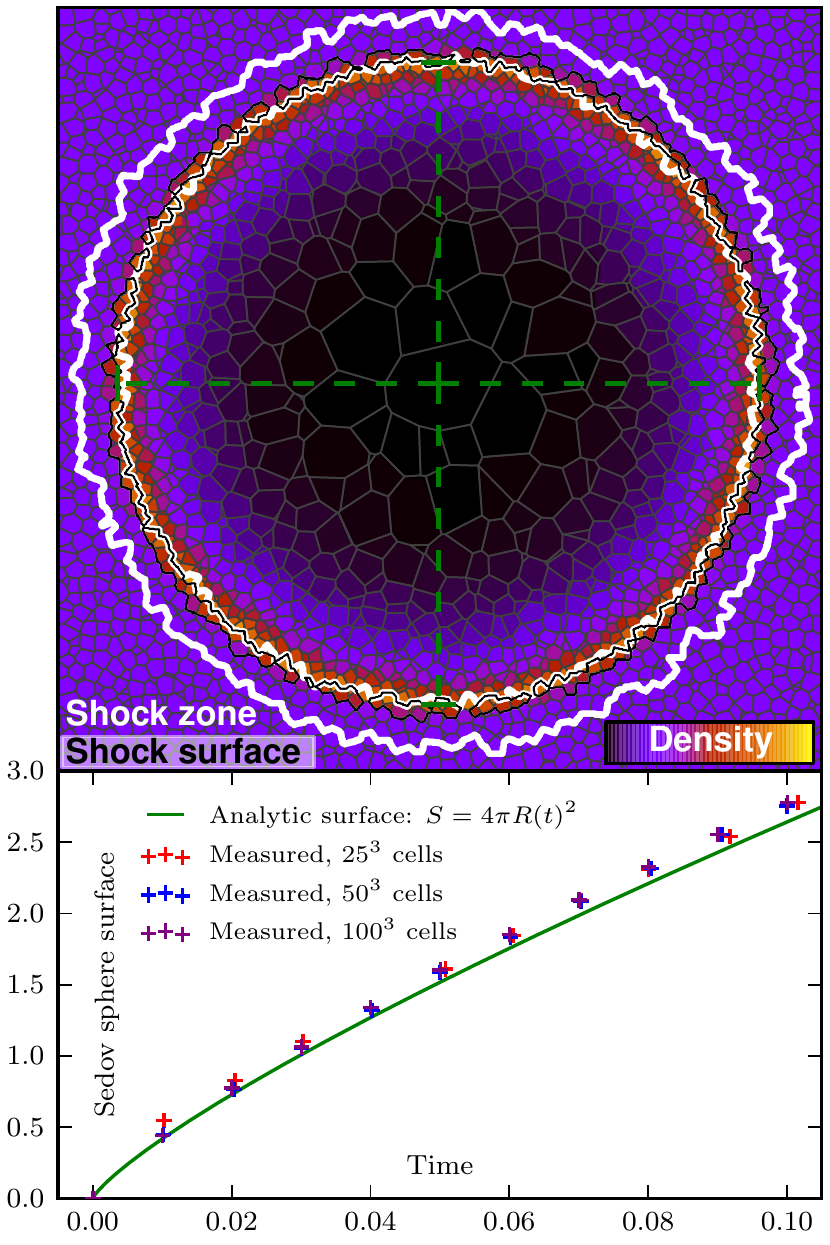}
 \caption
 { Top panel: cross-section of a three-dimensional Sedov blast
   wave simulation with $50^3$ cells at $t=0.08$ and energy $E=1$.
   The colour map encodes the density field of the fluid, and the
   green dashed cross marks the analytically calculated extent of the
   Sedov blast wave at this time. The cells inside the white contours
   belong to the identified `shock zone'; they fulfil the three
   criteria described in Section~\ref{sec:shock_zone}.  The black
   contours surround the cells that contain the reconstructed shock
   surface. These cells exhibit the minimum velocity divergence across
   the shock zone. Bottom panel: time evolution of the surface
   area of the spherical Sedov blast wave. We compare our approach to
   measure the shock surface in the test simulation (crosses) with the
   analytic evolution (solid line). For each cell in the shock
   surface, we assume an area contribution proportional to $V^{2/3}$
   with a prefactor calibrated with shock tube simulations.  }
\label{fig:sedov}
\end{figure}

The first part of our algorithm consists of a loop over all cells.  A
cell is flagged as being in the \textit{shock zone}, if the following
criteria are met:
\begin{align*}
&\text{(i)}\phantom{ii}\quad\mynabla\cdot\myvec{v}<0,\\
&\text{(ii)}\phantom{i}\quad\mynabla T\cdot \mynabla \rho > 0,\\
&\text{(iii)}\quad\Delta\log T\ge\log\left.\frac{T_2}{T_1}\right|_{\machnum=\machnum_{\text{min}}}\wedge\quad\Delta\log p\ge\log\left.\frac{p_2}{p_1}\right|_{\machnum=\machnum_{\text{min}}}.
\end{align*}
The first criterion is the standard compression criterion for shocks;
whenever a shock is present, this condition is true. It also in
principle filters tangential and contact discontinuities, however, this
is not effective in real-world numerical simulations.

The second criterion is constructed such that spurious shock
detections, potentially in a shear-flow or a cold front, are filtered
out.  Constant pressure implies that the density is inversely
proportional to temperature and therefore these variables increase in
opposite directions. At the same time, criterion (ii) holds in shocked
cells.

The third criterion is a numerical guard against detecting spurious
weak shocks. The first part of this protection mechanism introduces a
lower boundary for the temperature jump, as in \citet{RYU}.  The
second part demands a minimum pressure jump, which again discriminates
against tangential discontinuities and contacts. Note that this part
of criterion (iii) on its own may not be sufficient since
gravitationally compressed cells are also able to fulfil
it. $\Delta\log T$ and $\Delta\log p$ are calculated with the
temperature and pressure of neighbouring cells along the shock
direction.  The logarithm is taken such that the calculation can be
accomplished with a difference in order to avoid inaccurate divisions
in low temperature and pressure regimes.  In our analysis, the minimum
Mach number is set to $\machnum_{\text{min}}=1.3$ as in \citet{RYU}.  
We want to remark that the third criterion also rules out shocks with 
a slightly higher Mach number, since it is a local lower limit and the 
shock is broadened over a few cells. Note that we show in
Section~\ref{sec:shock_tube} that already $\machnum=1.5$ shocks are
fully captured.

In the following, we refer to the cells directly outside the shock
zone in the direction of the positive \mastervar gradient as
\textit{post-shock region}, while the corresponding cells in the
direction of the negative \mastervar gradient are referred to as
\textit{pre-shock region}.

\subsubsection{Shock surface}
\label{subsec:shock_surface}

After the determination of the shock zone, which has a typical
thickness of 3--4 cells, we proceed with the construction of a shock
surface consisting of a single layer of cells. For this purpose, rays
are sent from each cell of the shock zone in the direction of the
post-shock region (along the \mastervar gradient).  When the first cell
outside of the shock zone is reached, the post-shock \mastervar is
recorded and the ray direction is reversed in order to find the
pre-shock region. Furthermore, each ray stores the velocity
divergence of the cell from which it started. If a ray traverses a 
cell with a smaller divergence, the ray is discarded. For the rays
reaching the pre-shock region, the Mach number is calculated via the 
\mastervar jump of equation~(\ref{eq:t_jump}) and assigned to the original 
cell of the ray. We call these cells with minimum velocity 
divergence (i.e. maximum compression) across the shock zone the 
\textit{shock surface} cells. In this way, a Mach number is only 
calculated for cells in the shock surface. In the rare case that the 
direction of the temperature jump inferred from the pre- and post-shock 
temperatures is not consistent with the shock direction (given by the 
temperature gradient in the shocked cell), the detected feature is discarded.

In order to correctly treat overlapping shock zones of shocks
propagating in opposite directions, we calculate in each step along a
ray the scalar product of the shock direction of the original cell
with the shock direction of the current cell. If the product is
negative, the current temperature is recorded and the ray turns around
or stops, depending on whether it was heading for the 
post- or pre-shock region, respectively. With this approach 
we ensure that even when the shock zones of two different shocks 
overlap we are usually able to distinguish them and calculate their 
correct Mach numbers.

For the sake of bookkeeping simplicity in the distributed memory
parallelization of the algorithm, we send only one ray per shock zone
cell combined with reverting its direction once, instead of
simultaneously sending two separate rays in opposite directions.
Since the shock surface is very close to the post-shock region (see
Fig.~\ref{fig:sedov}), the maximum path a ray travels is only slightly
larger than the thickness of the shock zone.  Each ray starts at the
centre of mass of a cell and thereafter propagates from cell interface
to cell interface. The intersection between a ray and a Voronoi
interface is calculated analytically.  After all rays on the local {\small MPI}
task are propagated for one cell, the rays leaving the local domain
are communicated to the correct neighbouring task via a hypercube
communication scheme.

\subsubsection{Mach number calculation}
\label{sec:mach_number_calculation}

\begin{figure*}
\centering
\includegraphics{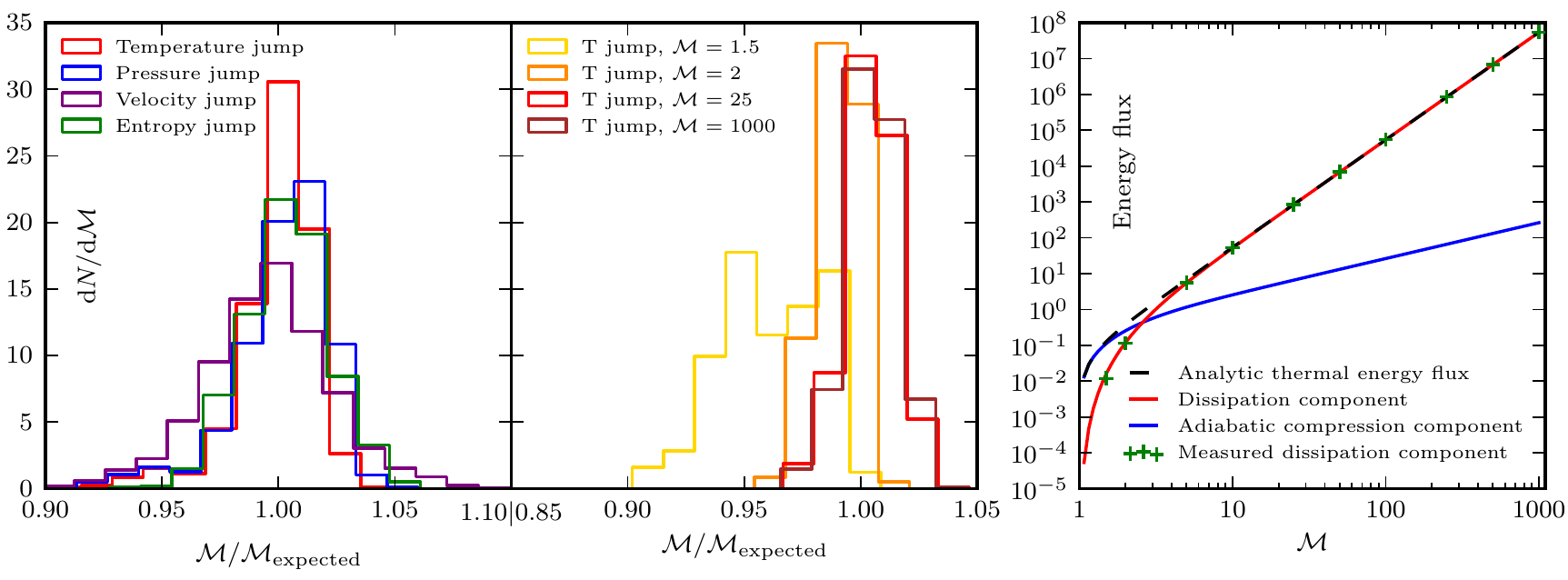}
\caption{Left-hand panel: number of shock surface cells $N$ per Mach
  number bin for ten different shock tube tests (see
  Table~\ref{tab:tube_inis}) for different calculation methods of the
  Mach number. We obtain the best results for the temperature jump of
  equation~(\ref{eq:t_jump}). Middle panel: Mach number
  distributions of single shock tubes obtained with the temperature
  jump method. Low Mach numbers such as $\machnum\simeq 1.5$ are
  slightly underestimated due to mild post-shock oscillations. This
  effect vanishes for Mach numbers $\machnum\ge 2$, where the correct
  value is found with an accuracy of 1 per cent.  In both the left-hand
  and middle panels, each histogram is normalized such that the area
  under the curve is unity. Right-hand panel: thermal energy fluxes
  in shock tubes, separately for the adiabatic and dissipative
  components. We compare the measurement using the temperature jump with
  the analytic solution, finding excellent agreement.  }
\label{fig:histo}
\end{figure*}

Given the pre- and post-shock values, the Mach number can in principle
be calculated with any of the equations ~\eqref{eq:rho_jump}--\eqref{eq:s_jump}.  
Note however that the Mach number calculation with
the entropy jump has to be accomplished with a numerical root finder,
for example a Newton--Raphson method. In Section~\ref{sec:shock_tube},
we investigate the quality of the practical results achieved
with each of these Mach number determination methods and conclude that
the \mastervar jump is best suited for the computation of the Mach
number in {\small AREPO}, see also Fig.~\ref{fig:histo}.

\subsection{Energy dissipation}

The thermal energy created at a shock can be expressed in terms 
of a generated thermal energy flux \citep{RYU}:  
\begin{align}
f_\text{th}=\left[e_2-e_1\left(\frac{\rho_2}{\rho_1}\right)^\gamma\right]v_2.
\end{align}
The indices 1 and 2 indicate the pre- and post-shock quantities,
respectively, and $e$ denotes the thermal energy per unit volume.
This flux can be expressed as a fraction of the incoming kinetic energy
flux $f_\Phi=\frac{1}{2}\rho_1(c_1\machnum)^3$:
\begin{align}
f_\text{th}=\delta(\machnum)f_\Phi.
\label{eq:f_thermal}
\end{align}
The thermalization efficiency $\delta(\machnum)$ can be calculated
from the Rankine--Hugoniot jump conditions \citep{KANG}, yielding
\begin{align}
\delta(\machnum)=\frac{2}{\gamma(\gamma-1)\machnum^2R}\left[\frac{2\gamma\machnum^2-(\gamma-1)}{(\gamma+1)}-R^\gamma\right],
\label{eq:delta_M}
\end{align}
where $R$ represents the density jump:
\begin{align}
R\equiv\frac{\rho_2}{\rho_1}=\frac{(\gamma+1)\machnum^2}{(\gamma-1)\machnum^2+2}.
\label{eq:R}
\end{align}
In our analysis, we use equation~\eqref{eq:f_thermal} for calculating
the generated thermal energy flux of a cell in the shock surface.
Given the area of the shock surface within a shocked cell (see
Section~\ref{sec:sedov_blast_wave}), we are then also able to
calculate the total dissipated energy per unit time.

Note that equations \eqref{eq:delta_M} and \eqref{eq:R}
describe the thermalization efficiency at shocks without considering
cosmic rays. The possibility of cosmic ray acceleration and the
corresponding efficiencies are addressed in Section~\ref{subsec:cosmic_ray_accel}.

\section{Validation}
\label{sec:validation}

\subsection{Sedov--Taylor blast wave} 
\label{sec:sedov_blast_wave}

We test the determination of the shock surface with simulations of
three-dimensional point explosions. We performed runs with $25^3$, $50^3$,
and $100^3$ cells. In order to obtain an unstructured Voronoi mesh free 
of any preferred directions for the initial conditions, we distribute
mesh-generating particles randomly in the unit box
$(x,y,z)\in[0,1]^3$. The mesh is then relaxed via Lloyd's algorithm
\citep{LLOYD} such that a glass-like configuration is obtained.  We
then set up the initial conditions as follows: the whole box is filled
with uniform gas of density $\rho_1=1$ and pressure $p=10^{-4}$, the
initial velocities are zero, and the adiabatic index is set to
$\gamma=5/3$. The energy $E=1$ is injected into a single central cell
of the grid.

We show a cross-section of the $50^3$ simulation at $t=0.08$ in the
top panel of Fig.~\ref{fig:sedov}.  At the corners of the box the
initial glass-like grid is still visible. Note however that the cross-section 
of a three-dimensional Voronoi grid is in general no longer a
Voronoi tessellation itself.  The colour of the cells represents the
density field of the fluid. The cells inside the white contours
constitute the identified shock zone. The shock surface consists of
the cells inside the black contour lines and features a position that
agrees well with the expected position (extent of the green cross).

Determining the correct shock surface area is important for
calculating the energy dissipation accurately.  We describe in
Section~\ref{sec:shock_tube} how we measure this area from the shock
surface cells.  In the bottom panel of Fig.~\ref{fig:sedov} we compare
the time evolution of the measured surface area of the Sedov shock
sphere with the analytic solution, which is given by $S(t)=4\pi
R^2(t)$ (green line), where $R(t)=\beta({Et^2}/{\rho_1})^{1/5}$
\citep{LL}.  The coefficient $\beta$ can be calculated numerically. We
obtain the value $\beta=1.152$ for $\gamma=5/3$ from the code provided
in \citet{KAMM}. Our measurement tracks the expected scaling well but
shows a small systematic overestimation of $\mytilde 5\%$. We suspect
the primary cause of the offset does not lie in the shock surface
estimation itself but rather appears because the simulated blast wave
is slightly ahead of the analytic solution due to low resolution
present at early times \citep{AREPO} in this self-similar problem.

\subsection{Shock tubes}

\label{sec:shock_tube}

We also checked the accuracy of the Mach number estimate for the
identified shock surface by performing numerous shock tube tests
\citep{SOD}. In view of our target applications, we chose to adopt a
three-dimensional box $(x,y,z)\in[0,100]\times[0,20]\times[0,20]$ in
all the tests. Again, a hydrodynamic glass-like initial grid is used
with $4\times10^4$ cells.  The gas has an adiabatic index of
$\gamma=5/3$, and the initial position of the discontinuity is prepared
at $x=50$.  The variables of the right state $(x>50)$ are set to
$p_\mathrm{r}=0.1$, $\rho_\mathrm{r}=0.125$, and $v_\mathrm{r}=0$.  
The density and the velocity
of the left state $(x<50)$ are $\rho_\mathrm{l}=1$ and $v_\mathrm{l}=0$,
respectively. Furthermore, we assign a pressure $p_\mathrm{l}$ to the left
state such that the shock has a specific Mach number, see
Table~\ref{tab:tube_inis}. The third column of the table shows the
simulation time at which the shock reaches $x=75$. We apply our shock
finder to the corresponding output file. Note that the shock finder in
this test, in contrast to the Sedov-Taylor blast wave, is also
confronted with rarefaction waves and contact discontinuities, which
obviously should not be mistaken as shock features by the shock finder.

\begin{table}
\begin{center}
\begin{tabular}{|c c c}
\toprule
$p_l$         & $\machnum$ & $t_{\text{end}}$\\ 
\midrule
0.81445190 & 1.5  & 14.43 \\ 
1.9083018 & 2.0 & 10.83 \\
15.357679 & 5.0 & 4.330 \\
63.498622 & 10.0 & 2.165 \\
400.51500 & 25.0 & 0.8660 \\
1604.1492 & 50.0 & 0.4330 \\
6418.6865 & 100.0 & 0.2165 \\
40120.448 & 250.0 & 0.08660 \\
160483.88 & 500.0 & 0.04330 \\
641937.62 & 1000.0 & 0.02165 \\
\bottomrule
\end{tabular}
\caption{Shock tube initial conditions. The pressure of the left state
  ($p_\mathrm{l}$) is varied such that the shock has a specific Mach number.
  The right-hand column indicates the time $t_{\rm end}$ when the shock has
  traversed three quarters of the tube, at which point we measure its
  strength with our shock finder implementation. 
\label{tab:tube_inis}}
\end{center}
\end{table}

The left-hand panel of Fig.~\ref{fig:histo} shows the quality of the Mach
number determination for all considered Mach number calculation
methods according to equations \eqref{eq:rho_jump}--\eqref{eq:s_jump}, 
except for the density jump method which is omitted
because it is not sensitive for high Mach numbers because
$\rho_2/\rho_1\rightarrow(\gamma+1)/(\gamma-1)$ for
$\machnum\rightarrow\infty$.  We note that in order to apply the
velocity jump method, the velocities have to be transformed into the
lab frame \citep{VAZZA}.

The overall best results with {\small AREPO} for the shock tube tests
are obtained with the temperature jump method according to
equation~\eqref{eq:t_jump}. It performs very well for Mach numbers
$\machnum\geq2$, as can be seen in the middle panel of
Fig.~\ref{fig:histo}. For small Mach numbers ($\machnum<2$), there
are mild post-shock oscillations which cause the temperature jump
method to underestimate the Mach number by a few percent.  This
systematic offset is present for all jump methods, unless the entropy
jump is used, which is not perturbed by these adiabatic oscillations.

For calculating the energy dissipation, the correct shock surface area
has to be determined.  The area contribution $S_i$ of a single cell to
the whole shock surface is expected to scale with its volume according
to $V^{2/3}_i$.  Furthermore, $S_i$ also depends on the shape of the
cell. Cells in a shock are compressed normal to the shock direction
and the degree of the compression depends on the strength of the
shock. We therefore make the ansatz $S_i=\alpha F_i^\beta V^{2/3}_i$,
where $F_i$ is the maximum face angle of the Voronoi cell, which
characterizes the shape of the cell. The definition of this quantity
has been introduced in \citet{VOGELSBERGER_STATISTICS} in the context
of a mesh regularization switch. We calibrate the constants $\alpha$
and $\beta$ with a least-square fit using the 10 shock tube problems
described above, where the total area of the shock surface is expected
to be equal to the cross-section of the tube ($S=400$). Our
calibration yields $\alpha=1.074$ and $\beta=0.4378$. By using these
values, we obtain for the mean shock surface area of the tubes
$\left<S \right>=396.35\pm2.45$, which is accurate to within $1\%$.
Note that in Fig.~\ref{fig:sedov} we have demonstrated that also
curved shock surfaces are measured to high accuracy.

With accurate Mach numbers combined with accurate shock surface areas in
each shocked cell, we are able to calculate the dissipated energy on a
cell-by-cell basis.  We show this explicitly in the right-hand panel of
Fig.~\ref{fig:histo}. The shock finder measures the dissipative
component of the thermal energy flux across a shock surface. The
second component contributing to the total thermal energy flux is
given by the adiabatic compression, which is present behind the
shock. As can be seen from the figure, the adiabatic thermal energy
flux is only relevant for small Mach numbers ($\machnum\leq5$).

\section{Shocks in non-radiative simulations}
\label{sec:non_radiative}

\subsection{Simulation set-up}

Besides full physics runs, the Illustris simulation suite
\citep{GENEL, ILLUSTRIS_NATURE, ILLUSTRIS_INTRO} contains also dark
matter only as well as non-radiative runs.  In this work we
investigate shocks in the non-radiative runs, which include dark
matter as well as gas, but no radiative cooling, star formation, and
feedback.  The cosmological parameters are consistent with the
9-year \textit{Wilkinson Microwave Anisotropy Probe} (\textit{WMAP}9) measurements
\citep{WMAP9}, and are given by $\Omega_{\mathrm{m}}=0.2726$,
$\Omega_{\Lambda}=0.7274$, $\Omega_{\mathrm{b}}=0.0456$, $\sigma_8=0.809$,
$n_\mathrm{s}=0.963$, and $H_0=100\,h\,\mathrm{km\,s^{-1}\,Mpc^{-1}}$ with
$h=0.704$. Two simulations with a resolution of $2\times455^3$ and
$2\times910^3$ were carried out in a periodic box having
$75\,h^{-1}\,\mathrm{Mpc}$ on a side, where the factor of $2$
indicates that the same number of gas and dark matter elements
were used. The adiabatic index of the gas is set to $\gamma=5/3$.
In the following, we refer to
these runs as Illustris-NR-3 and Illustris-NR-2, where
the latter is the one with the higher resolution.
The corresponding dark matter mass resolutions are 
$4.008\times10^8$ and $5.010\times10^7\,{\rm M}_\odot$;
the gas mass resolutions are kept fixed within a factor of 
$2$ at $8.052\times10^7$ and $1.007\times10^7\,{\rm M}_\odot$ 
(`target gas mass') by the quasi-Lagrangian nature as well as
the refinement and derefinement scheme of {\small AREPO}.
For redshifts $z>1$, the gravitational softening lengths of all mass 
components are equal and constant in comoving units,
growing in physical units to $2840$ and 
$1420\,\mathrm{pc}$ at $z=1$ for the NR-3 and NR-2 run,
respectively. For $z\le1$, the softening lengths of the 
baryonic mass components are kept fixed at these values.


\subsection{Shock finder assessment}

\begin{figure}
\centering
\includegraphics{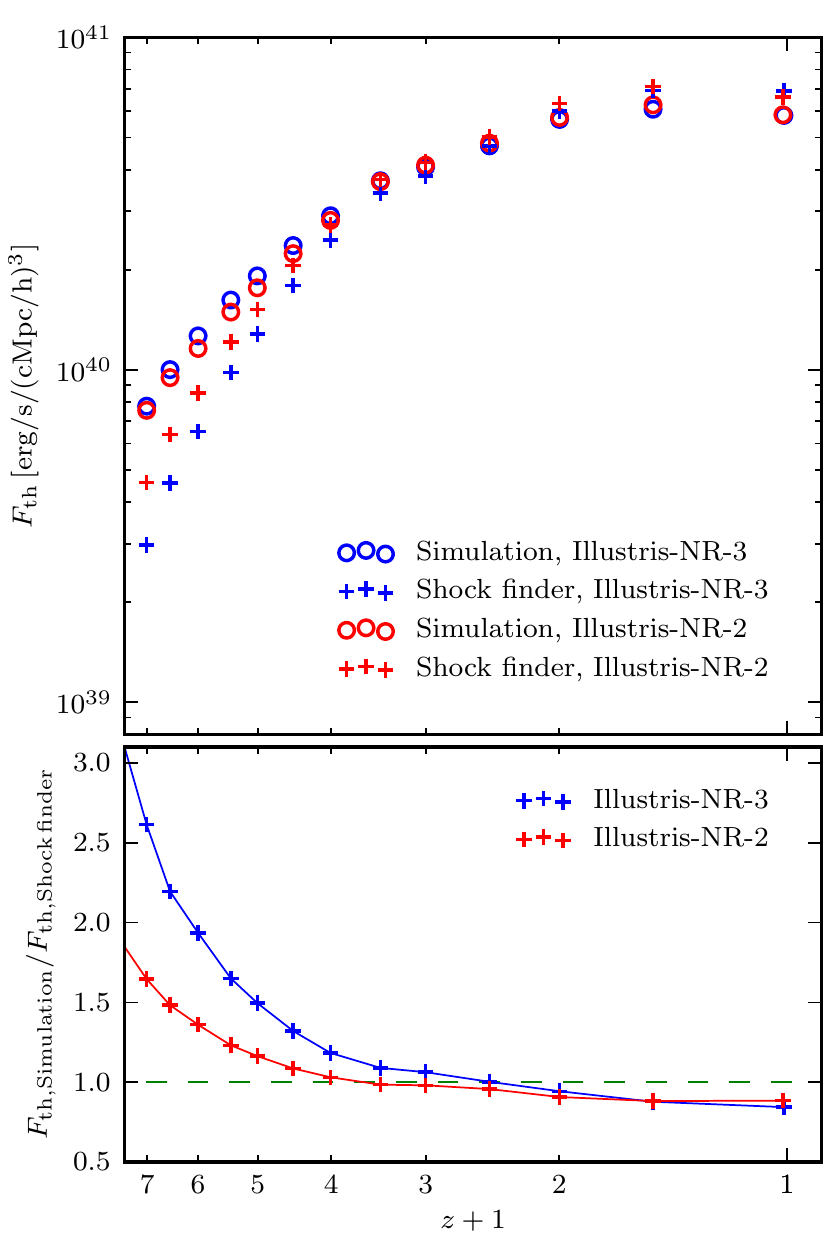}
 \caption
 {Comparison between the total energy dissipation rate found either
   with the shock finder, or inferred from two consecutive time steps
   of a non-radiative simulation.  The absolute values are shown in
   the top panel, the ratio is given in the bottom panel.  We find
   fairly good agreement for low redshifts, indicating that our shock
   finder properly accounts for all significant shock dissipation in
   the simulation. However, at high redshift not all the shock
   dissipation is recovered by the shock finder, an effect that
   diminishes greatly with better numerical resolution.  }
\label{fig:energy_comparison}
\end{figure}

First of all, we use the non-radiative runs for assessing the overall
quality of our shock finder.  To this end, we compare the total
dissipated energy per unit time obtained with the shock finder with
the dissipated energy measured between two consecutive time steps of
the simulation.  Under the assumption that the thermal energy during
one time step changes only due to dissipation and adiabatic
compression and expansion, we can write for every cell:
\begin{align}
u_2 m_2 - u_1 m_1 = \Delta E_{\text{diss}} + \left[u_1 \left(\frac{\rho_2}{\rho_1}\right)^{\gamma - 1} - u_1\right] m_1.
\label{eq:thermal_change}
\end{align}
Here $u$ denotes the thermal energy per unit mass, $m$ the mass,
$\rho$ the physical density, and the indices 1 and 2 correspond to
quantities before and after the time step, respectively.  We divide
equation \eqref{eq:thermal_change} by the time step size times the
comoving box volume and solve for $F_{\text{th, Simulation}}=\Delta
E_{\text{diss}}/(\Delta t V_\mathrm{c})$, the dissipated energy per time and
volume.  

The comparison with the shock finder measurement is shown in the top
panel of Fig.~\ref{fig:energy_comparison}. We find that our shock
finder recovers the full amount of dissipated energy for low redshifts
within $15\%$ accuracy. For very high redshifts, a progressively
larger deviation occurs and our shock detection results do not account
for all the dissipated energy any more. The origin of this difference
lies in the topology of early shocks, which are not yet pronounced and
resolved well at high redshift. Instead, they are rather scattered and
occupy a large fraction of the simulation volume.  As can be seen in
the bottom panel of Fig.~\ref{fig:energy_comparison}, the deviation
kicks in at progressively higher redshift for higher resolution
simulations. We hence conclude that our shock finder statistics has an
effective redshift completeness limit which depends on the resolution.
We can trust the shock detection results from $z=0$ up to $z\approx
4.0$ or $z\approx 5.0$ for the Illustris-NR-3 or Illustris-NR-2 runs, respectively.

\begin{figure*}
\centering
\includegraphics{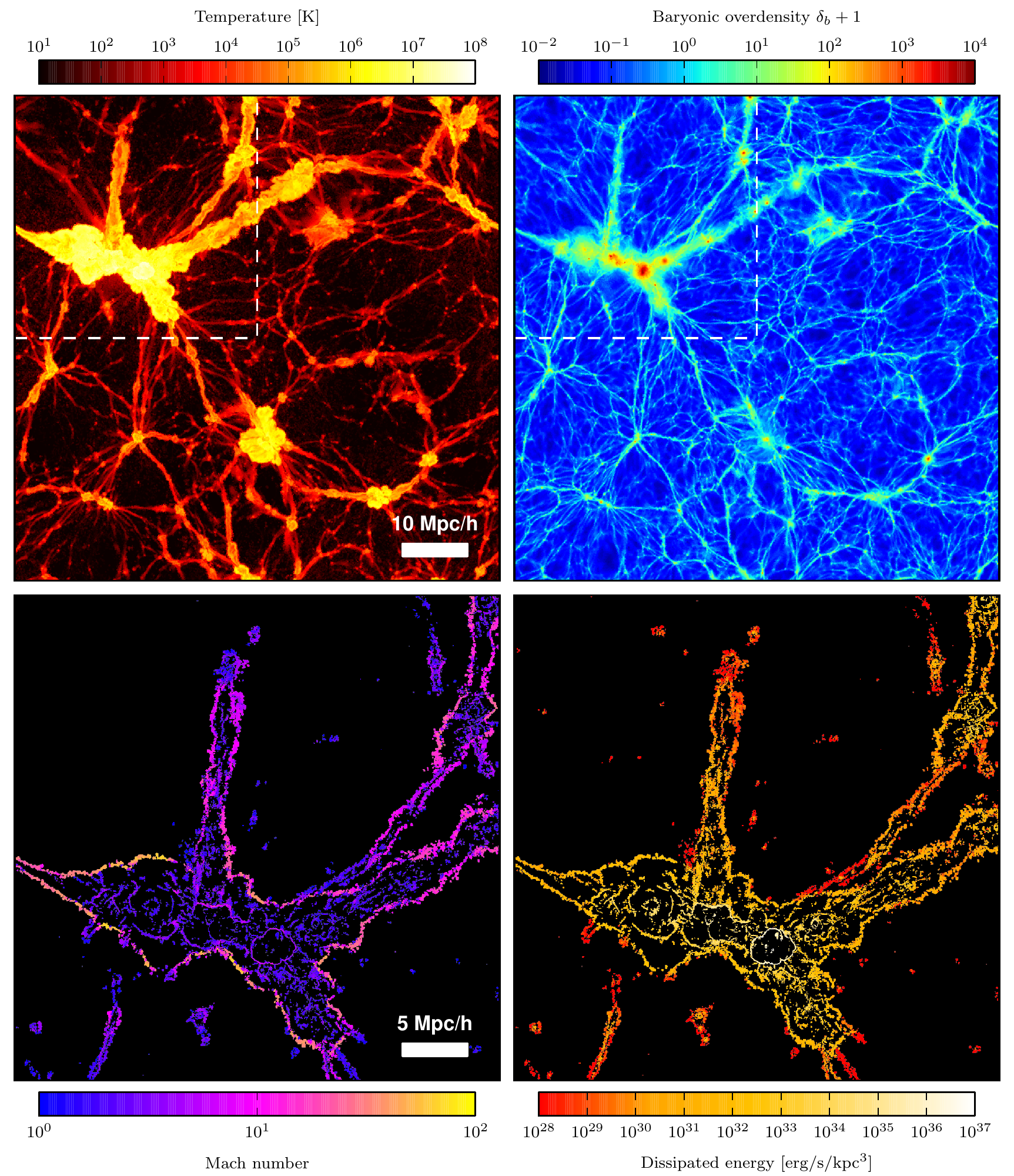}
\caption{Top panels: projections of the mass-weighted
  temperature and mean baryonic overdensity of the Illustris-NR-2 run at redshift $z=0$. The width and the
  height of the plots correspond to the full box size
  ($75\,h^{-1}\mathrm{Mpc}$). The projection in the $z$-direction has
  a depth of $150\,\mathrm{kpc}$ and is centred onto the biggest halo
  in the simulation. Bottom panels: Mach number field 
  weighted with the energy dissipation and mean energy
  dissipation rate density for the top left quarter of the box. Strong
  external shocks with Mach numbers up to $\machnum\sim 100$ onto the
  super-cluster are visible, as well as mostly weak shocks in the
  interior.  However, most of the energy gets dissipated internally
  due to the higher pre-shock density and temperature. Note that we do not 
  find many shocks inside the accretion shock onto the biggest halo, because
  here the gas motion is governed by subsonic turbulence, see also
  Fig.~\ref{fig:nr_halos}.}
\label{fig:nr}
\end{figure*}

\begin{figure*}
\centering
\includegraphics{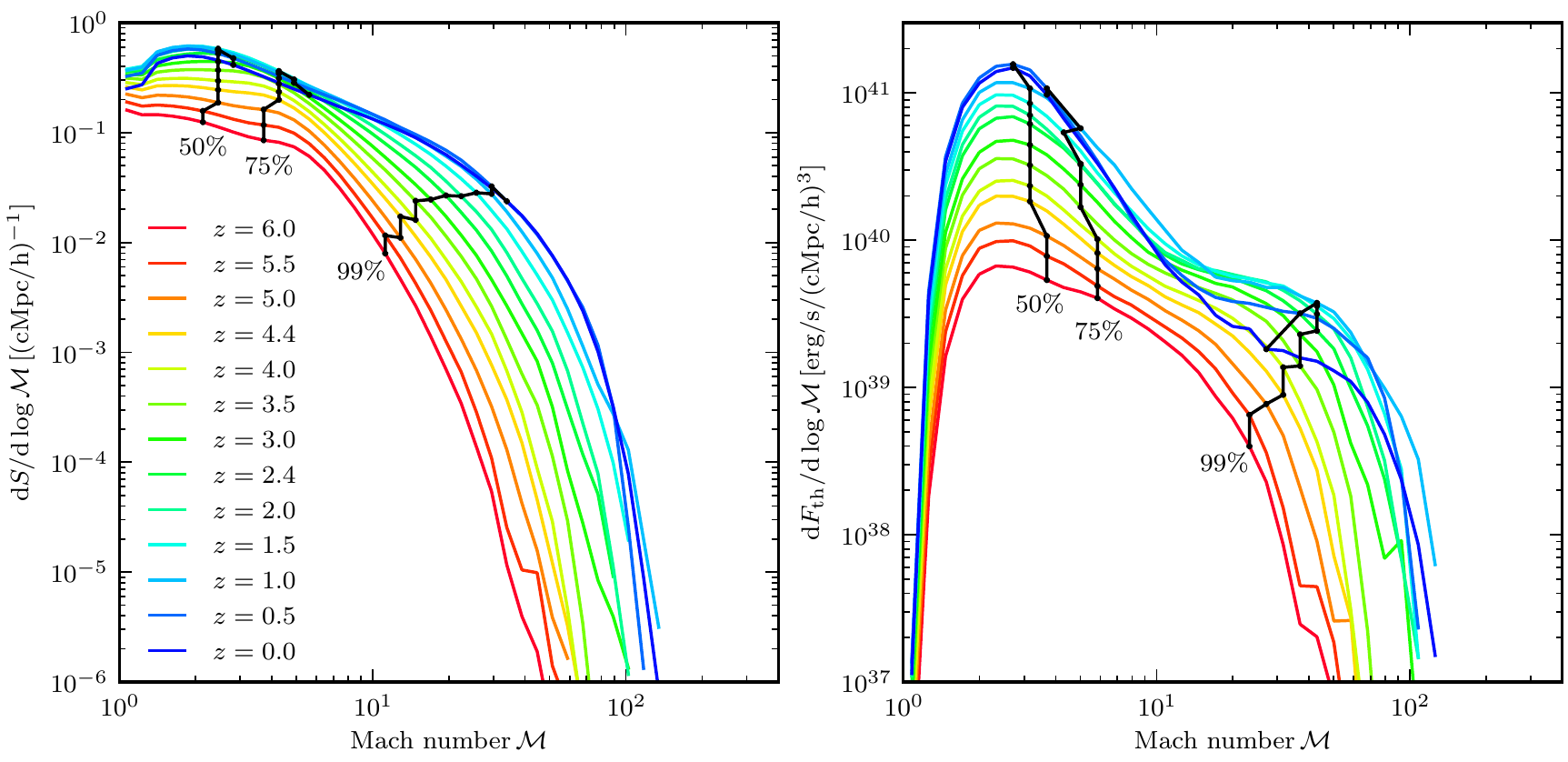}
\caption{Left-hand panel: differential shock surface area as a
  function of Mach number for different redshifts. The black lines
  indicate the Mach numbers up to which a specific fraction of the
  total surface is included.  Right-hand panel: distribution of the
  dissipated energy (generated thermal energy per time).  We find that
  shocks with $\machnum<6$ account for more than $75\%$ of all the
  shocks as well as $75\%$ of the total dissipated energy at all
  redshifts.  Compared to former studies, the peak of our energy
  dissipation distribution at $z=0$ is located at a considerable
  higher Mach number ($\machnum\approx 2.7$ instead of $\machnum\approx 2$).}
\label{fig:surface_energy}
\end{figure*}

\begin{figure*}
\centering
\includegraphics{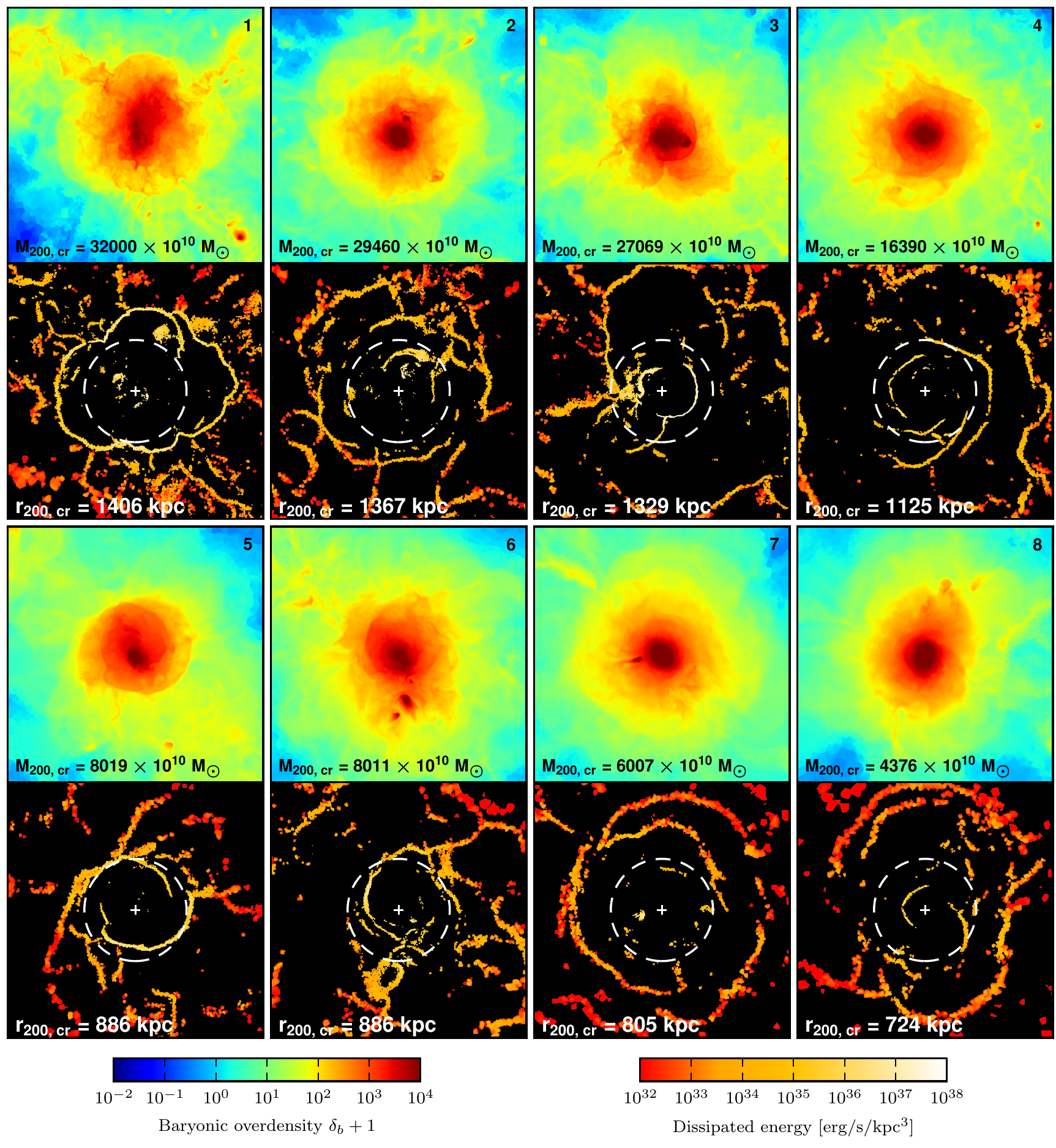}
\caption{Zoom projections with a width of $100\,\mathrm{kpc}$ centred
  on some of the biggest haloes in the Illustris-NR-2 simulation. The top panels
  show the baryonic overdensity while the bottom panels indicate the
  energy dissipation.  Accretion shocks onto the haloes can be found
  close to but outside of $r_{200,\,\mathrm{cr}}$ (white circles).
  Inside the accretion shocks prominent merger shocks are present. We
  do not find many shocks due to complex flow patterns within
  clusters, unlike reported by previous studies.  }
\label{fig:nr_halos}
\end{figure*}

\subsection{Reionization modelling}

The simulations Illustris-NR-3 and Illustris-NR-2 have no significant 
temperature floor and do not model cosmic reionization during their runtime.
However, it is important to account for the nearly uniform heating of the ambient
gas at the reionization redshift ($z\simeq 6-7$) in order to avoid
overestimating the Mach numbers of shocks from voids onto filaments at
late times.  For this purpose, we use a temperature floor of
$10^4\,\mathrm{K}$ for the shock finding carried out in
post-processing, the same procedure as used by \citet{RYU} and
\citet{SKILLMAN}. We can justify the simplicity of this approach by
the marginal contribution made by shocks with low pre-shock temperature
and density (voids onto filaments) to the dissipated energy, which is
the main quantity of interest in our analysis.  Nevertheless,
reionization in post-processing could of course be modelled in a more
sophisticated way, for example by a fitting function in the
density--temperature plane \citep{VAZZA}.

\subsection{General properties}

In Fig.~\ref{fig:nr} we present the state of the Illustris-NR-2 simulation at
redshift $z=0$. The projections were created by means of point
sampling, and the shown quantities are the mass-weighted temperature,
the mean baryonic overdensity, the Mach number weighted with the
dissipated energy, and the mean dissipated energy density.  The latter
two are displayed only for the top left-hand quarter of the former
projections, which show a supercluster including the biggest halo of
the simulation. This halo has a mass of
$M_{200,\,\mathrm{cr}}=3.2\times 10^{14}\,{\rm M}_\odot$,
corresponding to a virial radius of
$r_{200,\,\mathrm{cr}}=1.4\,\mathrm{Mpc}$.

\begin{figure*}
\centering
\includegraphics{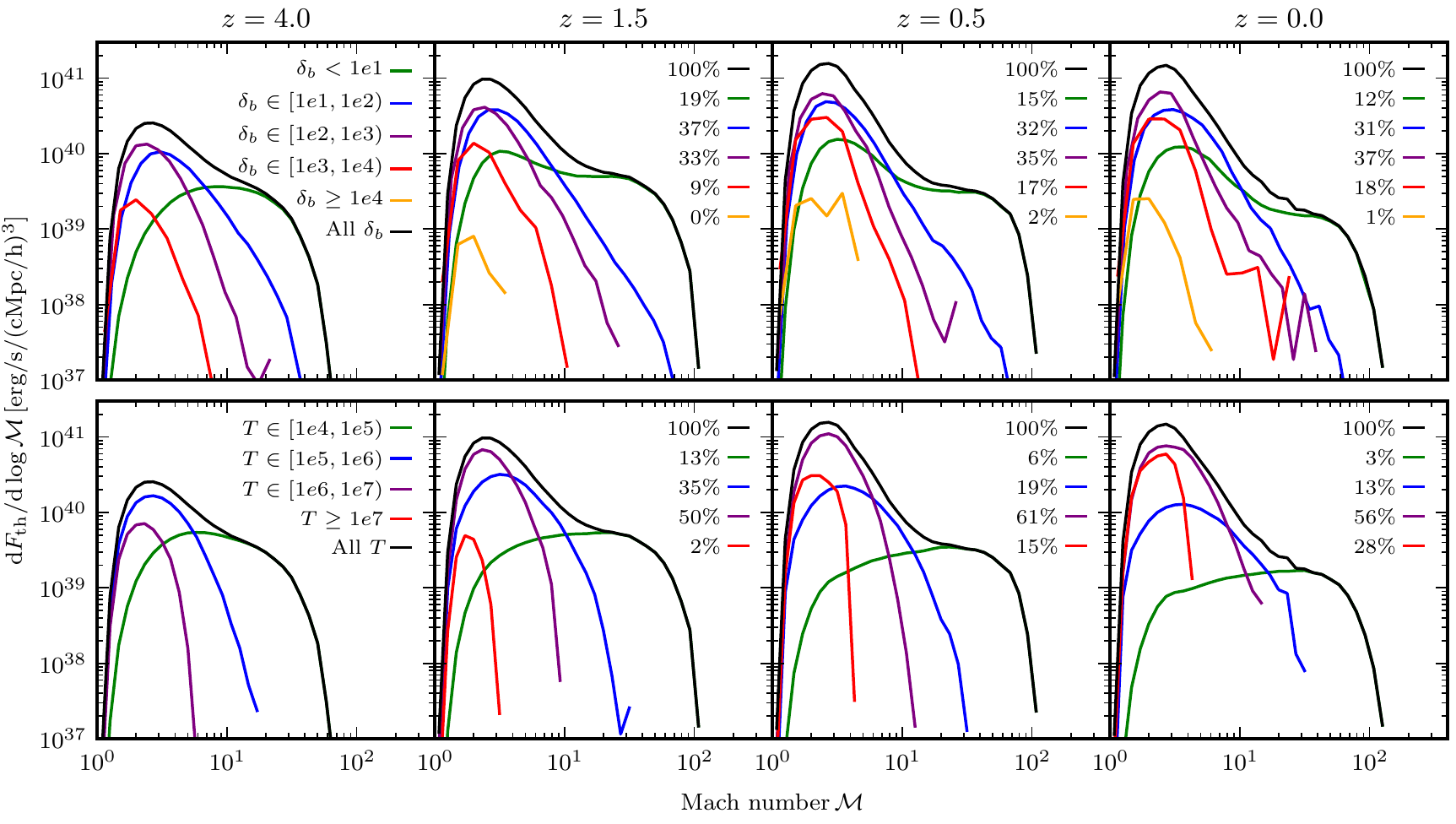}
\caption{Contribution of different baryonic pre-shock overdensities
  (top panels) and pre-shock temperatures (bottom panels) to the
  overall energy dissipation in the Illustris-NR-2 run.  We find that most of
  the energy dissipation at $z=0$ is due to shocks with pre-shock
  overdensities in the range $\delta_\mathrm{b} \in [10^1, 10^3)$ .
  Furthermore, around $70\%$ of the total dissipation is contributed
  by pre-shock gas with temperatures $T\in[10^5,10^7)$, of which most
  is located in the WHIM.  }
\label{fig:overs_temps}
\end{figure*}

Note that due to the temperature floor applied in post-processing only
the hottest filaments are detected by the shock finder and are hence
present in the bottom panels.  We can clearly observe the well-known
fact \citep{RYU} that high Mach numbers ($\machnum \sim 10 - 100$) are
generated at external shocks involving pristine pre-shock gas
($T_{\text{pre}}\lesssim10^4\,\mathrm{K}$), whereas gas previously
processed by internal shocks ($T_{\text{pre}}>10^4\,\mathrm{K}$)
experiences typically lower Mach numbers ($\machnum \lesssim 10
$). The density inside the supercluster exceeds the density of the
voids by several orders of magnitude and thus the energy dissipation
is most effective internally.  The highest dissipation rate in this
projection is present in the accretion shock onto the biggest cluster,
whereas we do not detect many shocks inside the accretion shock.

\subsection{Shock statistics}

Fig.~\ref{fig:surface_energy} quantifies the shock distribution and
the associated energy dissipation in the Illustris-NR-2 run. In the left-hand panel,
the differential shock surface area normalized by the simulation
volume is plotted as a function of the Mach number. We find redshift
independently that $50\%$ of the shocks have Mach numbers below $\machnum=3$, 
and $75\%$ below $\machnum=6$. Towards lower redshift the
cumulative area of shocks increases, especially for strong shocks. At
redshift $z=6$, shocks with a Mach number smaller than $\machnum=12$
account for $99\%$ of all the shocks, whereas at $z=0$ all the shocks
up to $\machnum=35$ make up $99\%$. At low redshift, the accretion
from previously unshocked gas onto hot filaments and cluster outskirts
provides Mach numbers up to $\machnum\approx100$.  At redshift $z=0$, the
total shock surface area reaches a value of $S=2.5\times
10^{-1}\,\mathrm{Mpc}^2/\,\mathrm{Mpc}^3$ (integral of the blue
curve).

The right-hand panel of Fig.~\ref{fig:surface_energy} shows the
differential thermal energy flux as a function of the Mach number. The
total dissipated energy increases with time up to $z=0.5$ and drops
thereafter slightly to a value of $2.3\times
10^{40}\,\mathrm{erg}\,\mathrm{s}^{-1}\,\mathrm{Mpc}^{-3}$ (see also
Fig.~\ref{fig:energy_comparison}, but be aware of the factor $h^3$).
The increase in time is expected due to an increasing number of shocks
and the ever higher pre-shock densities and temperatures found inside
structures. At low redshifts, this effect saturates and, furthermore,
dark energy slows structure growth and dilutes the pre-shock gas inside
voids, which leads to a drop of the thermal energy flux for high Mach
numbers. The latter observation has also been pointed out by
\citet{SKILLMAN}.  

We find that $50\%$ of the total energy dissipation occurs in shocks
with $\machnum<4$, and $75\%$ in shocks with $\machnum<6$. Mach numbers above
$\machnum>40$ do not contribute significantly to the dissipation.  We
find that the energy dissipation peaks for $z=6$ at $\machnum\approx
2.3$ and shifts towards $\machnum\approx 2.7$ for $z=0$. This peak position
at redshift zero differs considerably from the value $\machnum\approx
2$ found by different previous studies \citep{RYU, PFROMMER, SKILLMAN,
  VAZZA, VAZZACOMPARE}.  We believe that the origin of this
discrepancy lies in the improved methodology we adopt, and we will
elaborate more on this in Section~\ref{sec:variations}.

\begin{table}
\begin{center}
\begin{tabular}{c c c c c} 
\toprule
& \multicolumn{4}{c}{Temperature range [in K]}\\
&$[10^4,10^5)$ & $[10^5,10^6)$ &$[10^6,10^7)$ & $[10^7,\infty($ \\
\midrule
$\delta_\mathrm{b}<10^1$ &          $2\%$ & $4\%$ & $6\%$ & $0\%$\\
$\delta_\mathrm{b}\in[10^1,10^2)$ & $1\%$ & $6\%$ & $22\%$ & $3\%$\\
$\delta_\mathrm{b}\in[10^2,10^3)$ & $0\%$ & $2\%$ & $20\%$ & $15\%$\\
$\delta_\mathrm{b}\in[10^3,10^4)$ & $0\%$ & $1\%$ & $7\%$ & $10\%$\\
$\delta_\mathrm{b}\ge10^4$        & $0\%$ & $0\%$ & $1\%$ & $1\%$\\
\bottomrule
\end{tabular}
\label{tab:energy_diss}
\caption{Contributions to the total energy dissipation for different
  pre-shock temperature and baryonic overdensity ranges at redshift
  $z=0$.  We find that $\approx38\%$ of the energy gets dissipated by
  pre-shock gas of the WHIM ($T\in[10^5,10^7)$, $\delta_\mathrm{b}<10^2$), and
  $\approx 57\%$ of the thermalization happens in clusters and groups
  ($\delta_\mathrm{b}\ge10^2$). \label{tabshockdiss}}
\end{center}
\end{table}

Additional differences become apparent if we investigate the shock
locations.  In Fig.~\ref{fig:overs_temps} we show the distribution
of energy dissipation with respect to pre-shock densities (top panel)
and pre-shock temperatures (bottom panel). Note however that this plot
is not directly comparable to a similar investigation in
\citet{SKILLMAN} since their analysis is based on the inflowing
kinetic energy.  As expected, the relative contribution of shocks in
the densest regions increases with time while shocks in low density
gas become less important. At zero redshift, $68\%$ of the energy
dissipation is due to shocks with baryonic pre-shock overdensities
$10^1\le\delta_\mathrm{b}<10^3$.  On the other hand, only $19\%$ of the
dissipated energy heats cluster cores ($\delta_\mathrm{b} \ge 10^3$).

The temperature contributions show the typical bimodal distribution
consisting of external and internal shocks \citep{RYU}.  Previously
unshocked cold gas accretes in strong external shocks but accounts
only for a small amount of the dissipated energy.  On the other hand,
gas which gets shocked multiple times and is thus located inside
structures produces low Mach number shocks with high energy
dissipation.  At zero redshift, gas with pre-shock temperatures
$10^5\le T<10^7$ accounts for $69\%$ of the energy dissipation.
Moreover, in order to determine the contribution of the WHIM,
we examine the density of the gas in this
pre-shock temperature range and find that $\approx 54\%$ has a
baryonic overdensity below $\delta_\mathrm{b}=100$. A detailed break-up of the
dissipation rates in different bins of pre-shock density and temperature is
given in Table~\ref{tabshockdiss}.

We interpret our findings as follows. Almost $40\%$ of the
thermalization happens when gas from the WHIM gets shock heated,
whereas shocks inside the accretion shocks of clusters and groups
$(\delta_\mathrm{b}>100)$ account for $\approx60\%$ of the dissipation. Thus,
the relative importance of the WHIM is significantly higher than what
has been found in previous studies. Furthermore, the shocks we
identify in clusters and groups are prominent merger shocks rather
than stemming from halo-filling, complex flow patterns, as shown in
the next section.

\subsection{Galaxy cluster shocks}

Fig.~\ref{fig:nr_halos} shows zoom-projections of width
$100\,\mathrm{kpc}$ around different massive haloes of the Illustris-NR-2
simulation. The eight chosen haloes are sorted by mass, and we show the
baryonic overdensity as well as the energy dissipation by shocks. The
white circle indicates the virial radius $r_{200,\,\text{cr}}$. The
first halo is the biggest halo in the simulation and is also present
in Fig.~\ref{fig:nr}.  As can be seen in the gas projection, there are
turbulent motions inside the virial radius. However, we do not find
many shocks inside this region, which points towards predominantly
subsonic turbulence.

The third halo is in a state similar to the bullet cluster. The small
subcluster ploughs through the gas of the big halo and produces a bow
shock where a lot of energy is dissipated. The Mach number along the
bow shock is $2\lesssim\machnum\lesssim3.5$, comparable to the value
$\machnum=3\pm0.4$ measured for the bullet cluster system
\citep{MARKEVITCH2002, MARKEVITCH2006}.  The shocks in haloes four and
five form interesting spiral structures. These shocks might point
towards the following scenario: A minor-merger event triggers gas
sloshing and the formation of a spiral cold front \citep{ASCASIBAR,
  MARKEVITCH_COLD_FRONTS, ROEDIGER} which steepens to a shock while
propagating outwards.  A hint could also be the temperature map of
halo four, which indicates that the spiral shock structures fade into
contact discontinuities in the interior.  In halo six, there are fine
shock structures close to each other. 
In order to resolve these, it is
crucial to handle overlapping shock zones as described in
Section~\ref{subsec:shock_surface}.  The last two haloes are surrounded
by prominent accretion shocks. Furthermore, halo eight underwent a
merger event recently, as can be seen from the shock remnants inside
the virial radius lying opposite to each other. 
Several of such remnants have been observed in form of 
double radio relics, as reported for example in
\citet{ROTTGERING_1997}, \citet{BONAFEDE, BONAFEDE_2012}, and \citet{WEEREN_2009}. 
The accretion shocks
are located outside of the virial radius $r_{200,\,\mathrm{cr}}$. This
is expected since the kinetic and thermal pressure become roughly
equal around $r_{200,\,\mathrm{mean}}$ \citep{BATTAGLIA} which is
$\approx 1.5\,r_{200,\,\mathrm{cr}}$ at redshift $z=0$.

\begin{figure}
\centering
\includegraphics{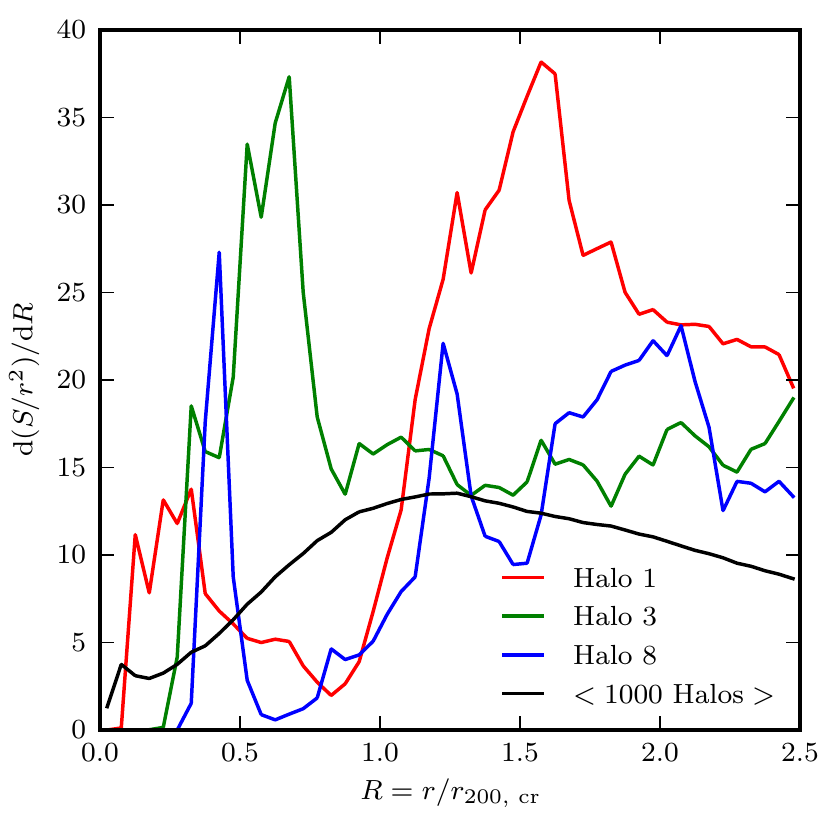}
 \caption
 {The radial shock surface distribution in several of the haloes of
   Fig.~\ref{fig:nr_halos} (coloured lines, as labelled) as well as the
   average distribution of the 1000 largest haloes in the
   Illustris-NR-2 simulation, stacked at redshift $z=0$. The average dissipation profile
   has a peak at around $R=1.3$ in units of the virial radius, which
   can be interpreted as the typical radius of the accretion shocks.
 }
\label{fig:radius_statistics}
\end{figure}

\begin{figure}
\centering
\includegraphics{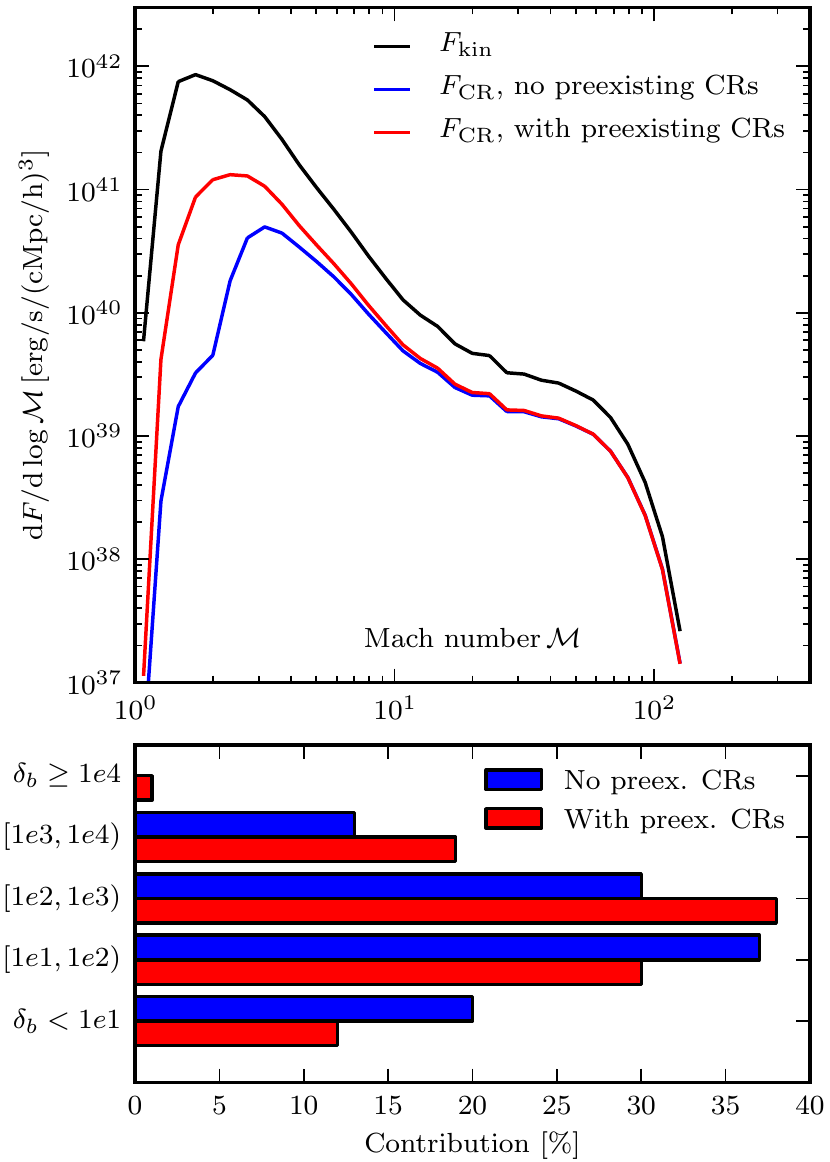}
 \caption
 {Top panel: Mach number dependent energy distribution for
   accelerating cosmic rays according to the DSA simulations of
   \citet{KANG}.  The black line shows the total kinetic energy
   processed by shocks per time and volume, while the red and blue
   curves give the fractions expected for particle acceleration with
   and without pre-existing cosmic rays, respectively. Bottom
   panel: contribution of different baryonic pre-shock overdensities
   to the total energy used for cosmic ray acceleration.}
\label{fig:cosmic_rays}
\end{figure}

\begin{figure}
\centering
\includegraphics{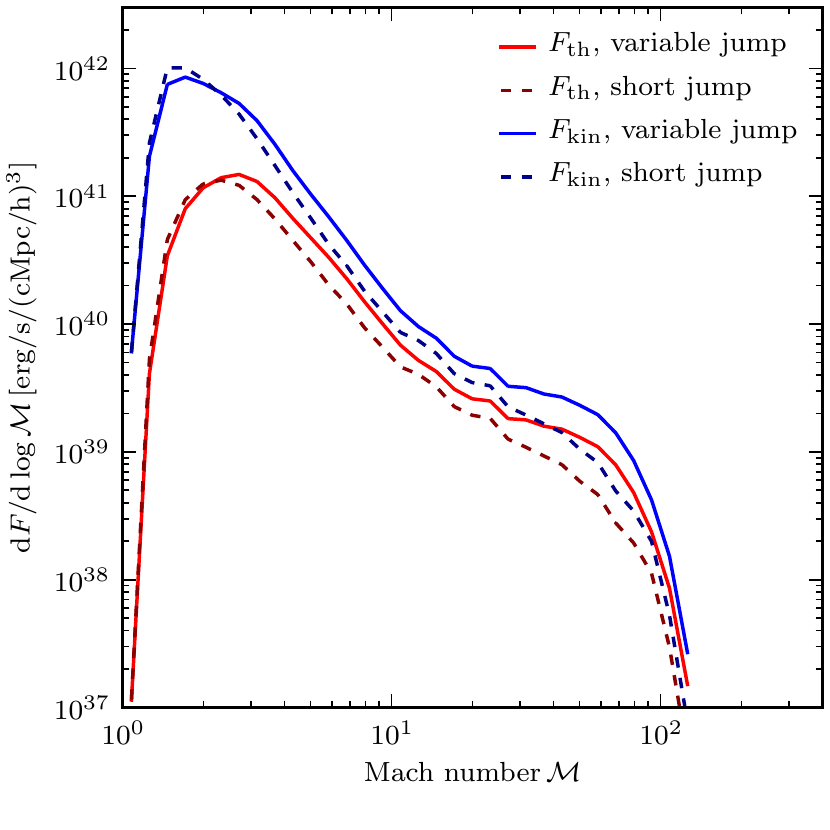}
\caption{Effect on the inflowing kinetic energy and
  dissipated energy when we calculate the Mach number based just on
  neighbouring cells (short jump) instead of using our standard
  implementation with a variable jump range. The restriction to short
  jumps shifts the distributions to slightly lower Mach numbers.}
\label{fig:variations1}
\end{figure}

In Fig.~\ref{fig:radius_statistics}, we present the radial shock
distribution for several of the haloes of Fig.~\ref{fig:nr_halos}. For
the biggest halo in the simulation (halo 1, red curve) the accretion
shock is located at around $1.6\,r_{200,\,\mathrm{cr}}$. The
green curve (halo 3) shows the bow shock of the bullet at
approximately $0.6\,r_{200,\,\mathrm{cr}}$, however, no single
radius for the accretion shock can be determined for this highly
dynamical system. In the blue curve (halo 8) three bumps can be
seen. The first one corresponds to the merger shocks inside the virial
radius ($R=0.4$). The two bumps outside the virial radius belong both
to the accretion shock and are located at the minor ($R=1.2$) and
major axis ($R\approx 1.9$) of the projected accretion shock
ellipsoid. In order to estimate an average radius for accretion
shocks, we stack the distributions of the 1000 largest haloes in the
simulation ($1.7\times 10^{12}\,{\rm M}_\odot\le {M}_{200,\text{
cr}}\le3.2\times 10^{14}\, {\rm M}_\odot$) and average them (black
curve). We find a signal with a broad peak at $R=1.3$, implying that
the accretion shocks at redshift $z=0$ can typically be found at
around $1.3\,r_{200,\,\mathrm{cr}}$.

\subsection{Cosmic ray acceleration}
\label{subsec:cosmic_ray_accel}

In this section, we discuss the role of the 
detected shocks in the Illustris-NR-2 run as cosmic ray sources. 
While our analysis is carried out
by post-processing simulation outputs,
cosmological simulations including cosmic rays and associated processes 
have also been presented, for example in 
\citet{MINIATI_2001_CR}, \citet{PFROMMER_2007_CR}, 
\citet{JUBELGAS_2008_CR}, and \citet{VAZZA_2012_CR}.

Cosmic rays get injected in collisionless cosmological shocks
by means of the DSA mechanism \citep[e.g.][]{BLANDFORD, MALKOVREVIEW},
also known as first order Fermi
acceleration. In this process, ions with high thermal energies can
diffuse upstream after crossing a shock and gain in a repetitive way
in multiple shock crossings more and more energy.  The cosmic ray
injection efficiency depends strongly on the Mach number as well as on
the level of Alfv\'{e}n turbulence, and is most efficient if the
magnetic field is parallel to the shock normal. Simulations of this
DSA scenario were carried out by \citet{KANG}, inferring upper
limits for the cosmic ray acceleration efficiency at specific Mach
numbers, in the case of a pre-existing as well as a non-pre-existing
cosmic ray population. Furthermore, fitting functions $\eta(M)$ are
provided for both cases and we use these for estimating the energy
flux $f_\text{CR}=\eta(M)f_\Phi$ transferred into cosmic rays, where
$f_\Phi$ is the processed kinetic energy flux.

The kinetic energy processed by shocks in the simulation as 
well as the estimated available energies for cosmic ray 
acceleration at redshift $z=0$ are presented in the top panel 
of Fig.~\ref{fig:cosmic_rays}.  The total inflowing kinetic energy 
amounts to $1.14\times10^{41}\,\mathrm{erg}\,\mathrm{s}^{-1}\,\mathrm{Mpc}^{-3}$.
Without a pre-existing cosmic ray population most of the acceleration
energy is available in shocks with Mach numbers $M\approx 3.2$ (blue
curve). Furthermore, if we compare the integrals of the distributions,
we estimate that $\approx6\%$ of the total kinetic energy processed by
shocks is transferred into cosmic rays. In the case of pre-existing cosmic
rays, which are generated in shocks at earlier times, the DSA
mechanism is more efficient, especially for low Mach numbers (red
curve). For this scenario, we find a peak position at $M\approx 2.5$
and a kinetic energy transfer to cosmic rays amounting to  $\approx18 \%$.

In order to determine the spatial origin of the cosmic rays, we show
the contributions of different baryonic overdensities to the total
cosmic ray energy in the bottom panel of
Fig.~\ref{fig:cosmic_rays}. For both scenarios (with and without
pre-existing cosmic rays) most of the energy transfer into cosmic rays
is located in regions with densities $10\le\delta_\mathrm{b}<10^3$ ($\approx
68\%$). Without a pre-existing population, only $43\%$ of the cosmic
ray energy is generated inside clusters and groups ($\delta_\mathrm{b}>100$). On the other
hand, pre-existing cosmic rays increase the efficiency for low Mach
number shocks, which are mainly present inside dense structures. In
this case we find that $58\%$ of the energy for cosmic ray
acceleration is provided by shocks inside clusters. We conclude that
cosmological shocks could produce a significant amount of cosmic rays.
Furthermore, the relevant shocks are shared in comparable proportions
among both, regions with overdensities above and below $\delta_\mathrm{b}=100$.

\begin{figure*}
\centering
\includegraphics{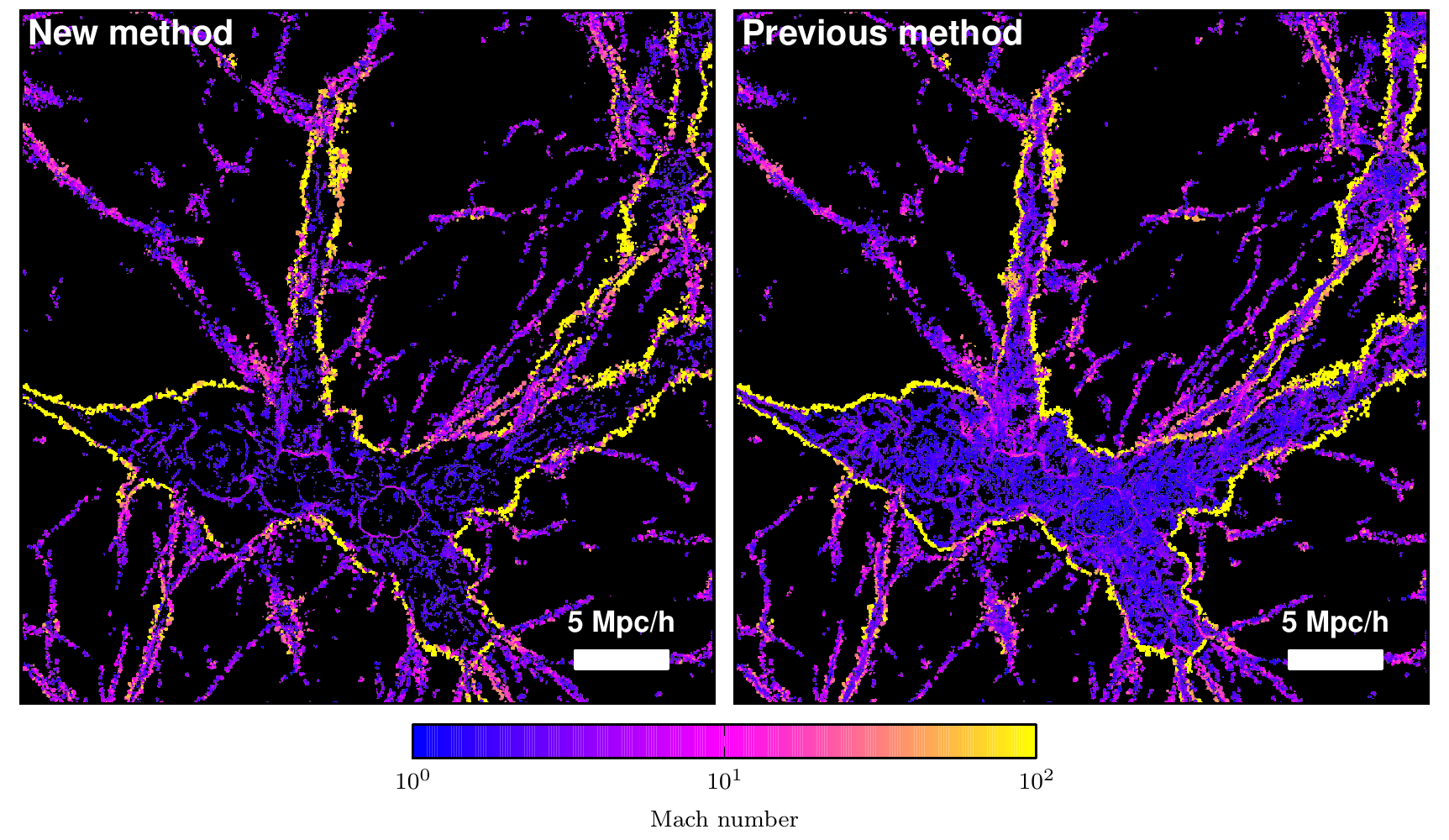}
 \caption
 {Demonstration of the difference between our default shock finder
   implementation (left-hand panel) and the old standard method (right-hand
   panel). With the latter implementation many more shocks are found
   inside the WHIM and inside clusters. These shocks are spurious as
   demonstrated in Fig.~\ref{fig:kh}. Depending on the applied
   evaluation method for the shock strength, some of the spurious
   shocks can be suppressed as shown in 
   Fig.~\ref{fig:variations2}. Nevertheless, if spurious shocks enter
   the analysis the relative contribution of clusters to the total
   dissipated energy is overestimated.  Note that for this
   comparison we do not adopt a temperature floor and thus find more
   shocks and higher Mach numbers compared to Fig.~\ref{fig:nr}.}
\label{fig:good_bad}
\end{figure*}

\section{Methodology variations}
\label{sec:variations}

In this section, we turn to an 
investigation of the origin of the discrepancies between our results and 
previous studies.  For this purpose, we in particular investigate the shock 
zone determination criteria as well as the Mach number calculation approach
for shock detection in the Illustris-NR-2 run.

For calculating the Mach number of shock surface cells, we normally
evaluate pre- and post-shock temperatures of cells just outside the
shock zone. An exception to this are overlapping shock zones, in which
case the pre-shock temperature is taken inside the combined shock zone
and between the shock surfaces. Note that with this procedure, Mach
numbers are calculated across a variable number of cells (variable
jump range), depending on the extent of the imprint of the shock on
the primitive variables. In
Fig.~\ref{fig:variations1} we show how the inflowing kinetic energy
distribution as well as the dissipated energy distribution changes
when we adopt a fixed jump range instead by calculating the Mach
number always from the directly adjacent cells of a shock centre
(short jump). Both distributions are shifted towards slightly lower
Mach numbers.  This finding is expected since a short jump does
not fully enclose the broadened shock, and should hence lead to a
smaller jump in the measured temperatures.  For the dissipated energy
distribution we find a peak shift from $\machnum\approx 2.7$ to $\approx 2.2$ 
as well as a reduction of the total dissipated energy
by $\approx 15 \%$ when the short jump is adopted.

In the next test we compare our default implementation of shock zone
finding to a method in which the requirement for a
minimum pressure jump ($\Delta\log
p\ge\log\left. p_2/p_1\right|_{\machnum=\machnum_{\text{min}}}$) is
abandoned, and the criterion $\mynabla T\cdot \mynabla \rho > 0$ is
replaced with $\mynabla T\cdot \mynabla S > 0$.  These relaxed
criteria have been frequently used in previous studies and only
recently \citet{HONG} suggested a replacement of the temperature--entropy criterion.

In Fig.~\ref{fig:good_bad}, we visualize the differences by applying
the two methods to the Illustris-NR-2 simulation, omitting a
temperature floor here for the purposes of this comparison test. With
the old standard shock finding method (right-hand panel), many more low
Mach number shocks are found inside the WHIM and inside galaxy
clusters. The additional shocks increase the total energy dissipation
and the relative contribution of clusters due to high pre-shock
densities and temperatures.

\begin{figure}
\centering
\includegraphics{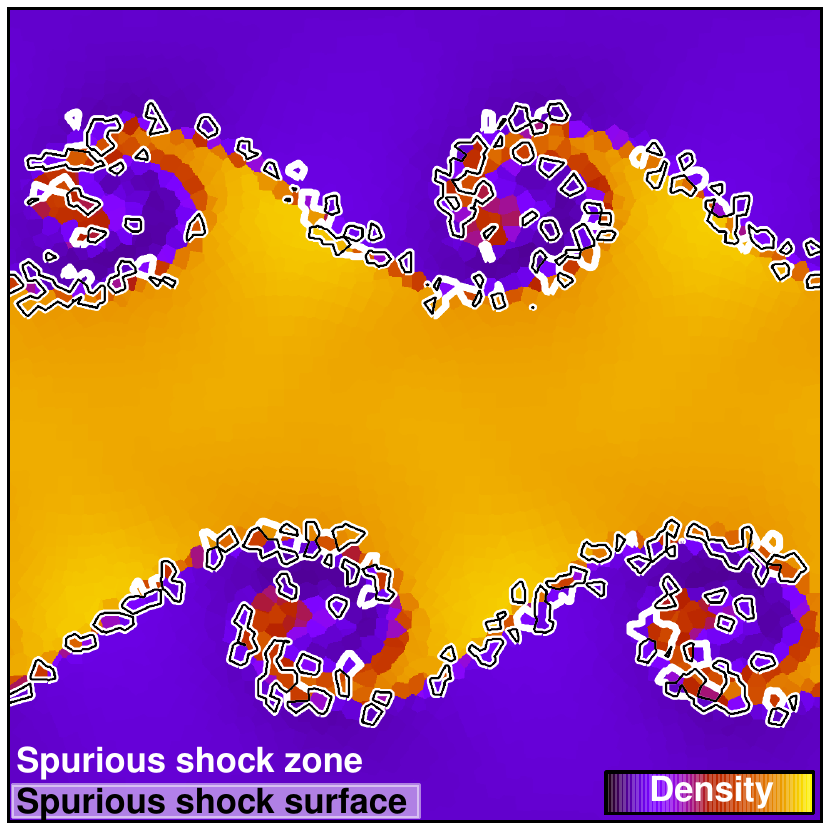}
\caption{Application of the previous standard shock finding method to
  a three-dimensional Kelvin--Helmholtz instability simulation. This
  method does not filter tangential and contact discontinuities and
  thus finds spurious shocks even though this subsonic problem is
  formally free of shocks. Our new implementation on the other hand does 
  not find shocked cells at all in the simulation volume, as desired. }
\label{fig:kh}
\end{figure}

We demonstrate in Fig.~\ref{fig:kh} that the additional shocks are
spurious by applying the old method to a three-dimensional
Kelvin--Helmholtz simulation in which no shocks are present.  However,
the old implementation is not able to filter tangential and contact
discontinuities reliably in the shock zone determination, and thus
false positive shocks are found along the density jump.

\begin{figure}
\centering
\includegraphics{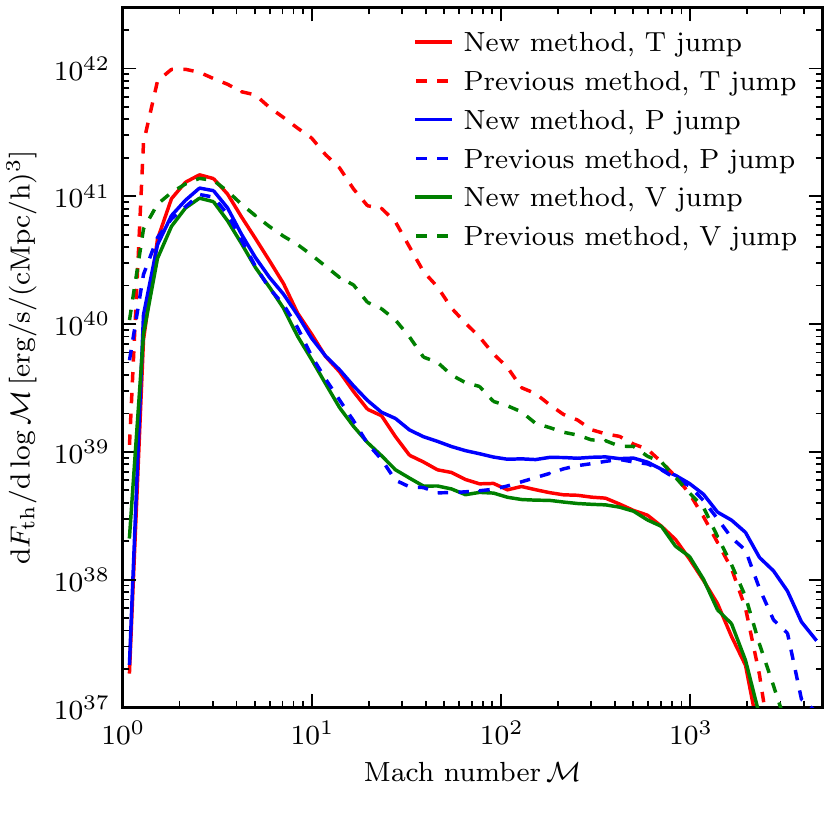}
\caption{Comparison of our new method to a previous standard
  method which has no minimum pressure jump requirement and
  uses $\mynabla T\cdot \mynabla S > 0$ instead of 
  $\mynabla T\cdot \mynabla \rho > 0$ as a shock zone criterion.
  The previous method does not
  filter tangential discontinuities in this case, see also
  Fig.~\ref{fig:kh}. We furthermore investigate the dependence of
  these methods on the Mach number estimation by using instead of 
  the temperature jump (our standard approach) also the pressure and
  velocity jumps, as labelled in the panel.}
\label{fig:variations2}
\end{figure}

Depending on the applied scheme for assessing the shock strength
(temperature jump, density jump, or velocity jump), some of the
spurious shock detections are suppressed by the consistency check that
requires a correct jump direction in the shock surface determination
(Section~\ref{subsec:shock_surface}). The energy dissipation inferred
for the different methods and different jump evaluations is presented
in Fig.~\ref{fig:variations2}. Note that the
reionization model through a temperature floor is disabled here, and
we therefore obtain a tail towards very high Mach numbers. It can be
seen that the old standard shock finding method is very sensitive to
the adopted jump quantity for inferring the Mach number, while our new
method is rather stable. There is good agreement between the
temperature jump and velocity jump methods, whereas the pressure jump
evaluation shifts the Mach numbers towards higher values for very
strong shocks.

The old standard method in combination with the temperature jump gives
clearly wrong results that are associated with an overestimate of the
energy dissipation, which can be clearly seen from the difference of
the dashed and the solid red line.  By using the velocity jump
instead, spurious shocks can be removed if the jump direction of the
normal velocities is not consistent with the shock direction
(temperature gradient). However, there remain then still a significant
number of cases with wrong shock detections, both with low and high
Mach numbers (green dashed line). Note that the pressure jump across
such spurious shocks is very small such that only a small Mach number
is assigned if the pressure discontinuity is used to measure the shock
strength. However, the distribution of energy dissipation at high Mach
numbers is also altered compared to our new method, and
perhaps surprisingly it is lower in this case (blue dashed line). The
origin of this effect lies in the modification of the Mach number of
real shocks by spurious shocks.  This happens whenever the shock zone
of a real shock gets modified (extended) by the shock zone of a
spuriously detected shock nearby.

We conclude that the use of a variable jump range gives a small
improvement compared to a fixed small jump range. It is yet more
important to properly filter against tangential and contact
discontinuities already in the shock zone determination to robustly
avoid spurious shock detections and distortions in the resulting shock
statistics.

\section{Summary}
\label{sec:summary}

We have implemented a parallel shock finder for the unstructured
moving-mesh code {\small AREPO}, based on ideas of previous work \citep{RYU,
 SKILLMAN, HONG} combined with new refinements.  Shocks are detected
in a two-step procedure. First, a broad shock zone is determined by
analysing local quantities of the Voronoi cells to identify regions of
compression.  In a second step, a shock surface is identified by
finding cells with the maximum compression across the shock zone,
followed by an estimate of the Mach number through measuring the
temperature jump across the shock zone.  In this way, the Mach number
is calculated over a variable number of cells which adjusts to the
numerical broadening of the particular shock.

Improvements to previous methods have been realized by handling
overlapping shock zones and by carefully filtering out tangential
discontinuities and contacts in the shock zone determination.  Such
discontinuities are abundantly present in cosmological simulations,
for example in the form of cold fronts and in shear flows that mix gas
of different specific entropy.  We robustly suppress spurious shock
detections by replacing the commonly used criterion 
$\mynabla T\cdot \mynabla S > 0$ with $\mynabla T\cdot\mynabla \rho > 0$ 
\citep[as also adopted in][]{HONG}, and additionally by requiring a suitable
minimum pressure jump threshold.

We have shown that shock finder results can be quite sensitive to
implementation details. For example, allowing for a variable number of
cells when evaluating the Mach number across shocks leads in general
to slightly higher (and more correct) values compared to adopting a
fixed number of cells. A still more important role is played by the
choice of fluid quantities selected for the Mach number
calculation. Especially when tangential and contact discontinuities
are not cleanly filtered out, an unfortunate choice can here produce
substantial distortions in the inferred shock statistics.

We have introduced a new test for assessing the overall performance of
a shock finder in which the energy dissipation inferred from the shock
detection is directly compared with the actual energy dissipation in a
non-radiative cosmological simulation between two consecutive time
steps. We find a good performance of our techniques for low redshifts
$z\lesssim 4.0\text{--}5.0$ in non-radiative cosmological simulations, where
we can identify all strong shocks reliably and account accurately and
in a numerically converged way for the bulk of significant dissipation
in shocks.  At zero redshift, the energy dissipation measured with our
shock finder for the adopted cosmological parameters is measured to be
$2.3\times 10^{40}\,\mathrm{erg}\, \mathrm{s}^{-1}\,
\mathrm{Mpc}^{-3}$.

Interestingly, we detected rich shock morphologies in the high-resolution 
non-radiative simulation Illustris-NR-2. In particular, high Mach
number accretion shocks onto filaments and cluster outskirts nicely
trace the cosmic web. Accretion shocks onto galaxy clusters dissipate
a lot of energy, while merger shocks inside the clusters give hints
about their recent formation history. We note that the merger shocks
appear as rather prominent and distinct features, whereas we do not
find complex flow shock patterns inside the cluster accretion shocks,
suggesting that there the gas dynamics is mostly characterized by
subsonic turbulence \citep[which we expect to be well captured by {\small
  AREPO};][]{BAUER}.

With our improved methodology, we find quantitatively revised results
for the shock dissipation statistics in non-radiative cosmological
simulations. Most of the thermalization happens in shocks with Mach
numbers around $\machnum\approx2.7$. Moreover, almost $40\%$ is
contributed by shocks in the WHIM and $\approx 60\%$ by shocks in
clusters and groups.  Compared to previous studies, these findings
correspond to a shift in the energy dissipation spectrum towards
higher Mach numbers and towards structures with lower densities.
Also, we have found $R=1.3\, r_{200,\,\mathrm{cr}}$ as a typical
radius for accretion shocks onto galaxy clusters at redshift $z=0$,
based on identifying a peak when stacking the radial shock dissipation
profiles of $1000$ haloes of the Illustris-NR-2 simulation. Consequently, the
accretion shock is expected to typically lie slightly outside the
virial radius, a finding which is consistent with
\citet{BATTAGLIA}. We note however that the accretion shock position
shows a high degree of temporal variability in any given halo.

Finally, we also investigated the expected energy transfer to cosmic
rays in the identified large-scale structure shocks if acceleration
efficiencies derived from DSA plasma
simulations are adopted \citep{KANG}. These simulations are set up
with a magnetic field parallel to the shock normal direction and
provide therefore upper limits for the acceleration efficiency.  We
obtain at redshift $z=0$ an average cosmic ray energy injection rate of
$7.0\times 10^{39}\,\mathrm{erg} \, \mathrm{s}^{-1}\,
\mathrm{Mpc}^{-3}$ in the case of non-pre-existing cosmic ray
populations, and a considerably larger value in the case of pre-existing
cosmic rays.  Considering these numbers, it is quite plausible that even for
random magnetic field orientations a dynamically important cosmic ray
population is produced in these shocks.  Furthermore, we found that
gas with pre-shock overdensities above and below $\delta_\mathrm{b}=100$
contribute roughly equally to the energy transfer into cosmic rays.

In future work, it will be interesting to couple the shock finder to
hydrodynamical simulations that take cosmic rays self-consistently
into account. Also, it should be interesting to contrast the results
obtained here for non-radiative simulations with an analysis of full
physics simulations of galaxy formation that include radiative cooling
and heating mechanisms, as well as prescriptions for star formation,
stellar evolution, black hole growth, and associated feedback
processes.  These simulations feature interesting additional shocks,
for example due to strong feedback-driven outflows. In a companion
paper (in preparation), we will present an analysis of the
corresponding shocks in the recent Illustris simulation
\citep{ILLUSTRIS_NATURE}.

\section*{Acknowledgements}

We thank Christoph Pfrommer, R\"{u}diger Pakmor, Andreas Bauer, and Shy Genel
for very helpful discussions.  The authors acknowledge financial support
through subproject EXAMAG of the Priority Programme 1648 `SPPEXA' of
the German Science Foundation, and through the European Research
Council through ERC-StG grant EXAGAL-308037.

\bibliography{literature} 
\bibliographystyle{mn2e}

\label{lastpage}

\end{document}